\documentclass[12pt,preprint]{aastex}

\usepackage{epsfig}
\usepackage{graphicx}
\usepackage{epstopdf}
\usepackage{psfig}
\usepackage{longtable}

\usepackage{natbib}
\begin{document}

\title{Trigonometric Parallaxes for 1,507 Nearby Mid-to-late M-dwarfs}
\author{Jason A. Dittmann$^{1}$, Jonathan M. Irwin$^{1}$, David Charbonneau$^{1}$, Zachory K. Berta-Thompson$^{2}$}
\affil{[1] Harvard-Smithsonian Center for Astrophysics, 60 Garden St., Cambridge, MA, 02138} 
\affil{[2] Massachusetts Institute of Technology, Cambridge, MA 02139}

\begin{abstract}
The MEarth survey is a search for small rocky planets around the smallest, nearest stars to the Sun as identified by high proper motion with red colors. We augmented our planetary search time series with lower cadence astrometric imaging and obtained two million images of approximately 1800 stars suspected to be mid-to-late M dwarfs. We fit an astrometric model to MEarth's images for 1507 stars and obtained trigonometric distance measurements to each star with an average precision of $5$ milliarcseconds. Our measurements, combined with the 2MASS photometry, allowed us to obtain an absolute $K_s$ magnitude for each star. In turn, this allows us to better estimate the stellar parameters than those obtained with photometric estimates alone and to better prioritize the targets chosen to monitor at high cadence for planetary transits. The MEarth sample is mostly complete out to a distance of 25 parsecs for stars of type M5.5V and earlier, and mostly complete for later type stars out to 20 parsecs. We find eight stars that are within ten parsecs of the Sun for which there did not exist a published trigonometric parallax distance estimate. We release with this work a catalog of the trigonometric parallax measurements for 1,507 mid-to-late M-dwarfs, as well as new estimates of their masses and radii.

\end{abstract}

\section{Introduction}
One of the goals of stellar research is to obtain the complete census of stars within the solar neighborhood out to a specified distance. A complete volume-limited sample will inform us about the stellar mass function, stellar formation, and the kinematics of the galaxy as well as the nearby stellar clusters for which we can identify members. Low mass stars vastly outnumber higher mass stars like the Sun, and so the main difficulty in constructing a volume limited sample is in identifying nearby, low mass, low luminosity objects and obtaining accurate distances to them. Apparent magnitude is a poor indicator of proximity; only a couple dozen of the several thousand stars visible by the naked eye are within 10 parsecs of the Sun \citep{luyten_obituary}. Currently, the most straightforward method to identify potentially nearby stars is from proper motion surveys, as a star with a high angular velocity is more likely to be nearby to the Sun. Proper motion surveys have been conducted for many decades, as the measurement is relatively simple to make, requiring only images of the same field separated by a length of time. Proper motion surveys continually improve, as longer time base lines increase the accuracy of the measurement. A uniform census of nearby stars allows for characterizing the relative occurrence rates of different types of stars, and to plot the relationship between intrinsic properties of those stars, including absolute magnitude and color.

Early attempts at conducting a large survey for high proper motion stars began in the early 20th century with work by \citet{van_maanen_1915}, who compiled a list of stars known at that time whose proper motion exceeded 0.50 arcseconds per year. This list was subsequently expanded by \citet{wolf1919} and \citet{Ross39}, pushing the limit to $0.20$ arcseconds per year. As these surveys progressed, it became apparent that there existed a large population of high-proper motion, low luminosity objects that were previously undetected due to their faintness. Due to the relatively large numbers of these objects, obtaining a volume-limited sample of high proper motion stars required deep exposures of the entire sky, combined with long time baselines. Such a survey was completed over several years by various groups, most notably from Lowell Observatory \citep{giclas_71, giclas_78}, which consists of 11,749 stars, and the New Luyten Catalogue of Stars with Proper Motions Larger than Two Tenths of an Arcsecond (the NLTT catalog; \citealt{nltt_catalog}), which contains 58,845 objects. More recently, \citet{Lepine_catalog} compiled a list of 61,977 stars in the northern hemisphere with proper motions larger than 0.15 arcseconds per year, identifying over 90\% of those stars down to a limiting magnitude of $V \approx 19.0$, excepting the galactic plane. A subset of this catalog identified in \citet{Lepine_33pc_sample} identifies those stars that are likely within 33 pc of the Sun, in order to provide the basis for a volume limited sample for further study and characterization, including spectral typing and obtaining direct distance estimates to these stars.

Of the several different methods that can be used to obtain distances to stars, trigonometric parallax is the most accurate. Photometric and spectroscopic parallax methods, in contrast, assume that the star is single and that the main sequence is single valued with luminosity as a function of effective temperature. When estimating the distances to stars through these methods, one might introduce systematic errors depending on the quality of the data and the models that are assumed. 

The first detection of a stellar parallax was for 61 Cyg by \citet{bessel}. A catalog of 248 stars with parallaxes determined from plates taken by Prof. Donner at Helsingsfors from 1891-1892 was released by \citet{Kapteyn1900}. Such long delays in parallax studies became common, as systematic effects between observers and observatories often resulted in widely different claimed values for the parallax of individual stars. A compilation by \citet{french_catalog} of trigonometric parallaxes known at that time lists 23 individual measurements of the parallax for 61 Cyg, ranging from $177$ milliarcseconds to $559$ milliarcseconds. The currently accepted value is $285.4$ milliarcseconds with an uncertainty of 0.8 milliarcseconds \citep{Hipparcos_original}. 

Great effort was made to reduce the systematic error in parallax work by Schlesinger, beginning in 1903 with the Yerkes 40-inch refractor \citep{Schlesinger04}. He utilized newly designed plates to spread blue light out over enough area to be effectively undetectable, in order to avoid systematic errors in the derived positions associated with refraction in the plate material \citep{Schlesinger_films, Schlesinger_glass}. These efforts began yielding fruit (see \citet{Yerkes1, Yerkes2}, and companion papers), ultimately producing a catalog of 1870 parallaxes \citep{1924_cat}. These techniques were widely adopted among many observatories, with new measurements being published in batches of dozens to hundreds. The total number of stars with parallaxes grew to 5822 with the compilation of the first Yale Parallax Catalog \citep{1952_cat}.

Astrometry programs have since conducted larger surveys for dimmer objects. Today, the two major databases of stellar parallaxes come from the General Catalog of Trigonometric Stellar Parallaxes, Fourth Edition (the Yale Parallax Catalog, \citealt{YPC}) and the Hipparcos mission \citep{Hipparcos_original, Hipparcos_reduction_1, Hipparcos_reduction_2}, which combined have distance determinations to approximately $120,000$ stars. Precisions in the Yale Parallax catalog range from a few milliarcseconds to 20 milliarcseconds, while the Hipparcos catalog routinely obtains precisions of several tenths of milliarcseconds for stars brighter than $V$ = 8. This brightness cutoff of parallax programs means that low mass objects such as most M-dwarfs and brown dwarfs are systematically underrepresented in these surveys and almost completely absent from the Hipparcos sample.

Numerous trigonometric parallax programs are currently underway to identify and determine distances to the nearby missing objects, consisting mainly of low mass stars and brown dwarfs. As these objects are cool and dim in the optical, parallax programs using infrared instruments have proved successful \citep{trent,USNO_IR_parallax}. Additionally, the Research Consortium on Nearby Stars (RECONS) survey\footnote{RECONS (http://www.chara.gsu.edu/RECONS/) has graciously made their results easily accessible, and we have used their recent parallax work extensively when comparing to our own.}, a long standing parallax program at CTIO, has provided a wealth of parallaxes from nearby white dwarfs, brown dwarfs, and late M dwarfs, currently with 30 papers identifying and characterizing new stars in the solar neighborhood, many of which lie within 10 parsecs (see \citealt{ctio} and references therein). RECONS has utilized astrometric, photometric, and spectroscopic observing techniques in order to discover and characterize these systems. Limited compilations of large numbers of measurements of these objects are also given by \citet{LandG11}. \citet{LandG11} aim to provide a candidate list of nearby M-dwarfs with high proper motion for further studies by other groups. Understanding these stars and their companions requires a reliable distance estimate, in order to estimate their intrinsic luminosities, radii, and other physical parameters. 

The small sizes of M dwarfs make them excellent targets for ground based searches for transiting small planets. This contributed to a renewed interest in these objects, and obtaining a volume limited sample, and accurate distances, are important scientific goals. MEarth is an ongoing photometric survey of nearby ($D \lesssim 33$ pc) mid-to-late M dwarfs, designed to be sensitive to planets around these stars as small as $2 R_\earth$ and with periods extending into the habitable zone \citep{Nutzman}. The advantages of conducting a transiting planet search around these stars include being sensitive to rock and ice planets from the ground, the shorter period of a habitable-zone planet, and the accessibility of the atmospheres of these planets with current or next-generation instrumentation (eg - JWST and the next generation of large ground-based telescopes). The first planet discovered by MEarth, the super-Earth GJ 1214b \citep{gj1214b}, has been the subject of an intense series of observations meant to measure its atmosphere from the optical (e.g. \citealt{gj1214_rayleigh, gj1214_bean, Murgas}), to the infrared (e.g. \citealt{gj1214_hubble, gj1214_bean, Crossfield_gj1214_spectrum, gj1214_croll, Desert_atmo, Fraine}). 

The characterization of GJ 1214b was aided largely by the amount of information already known about the star prior to the discovery of GJ 1214b, including its trigonometric parallax. This parallax allowed a decent estimate of the true size of the star and hence the planet. The relatively poor precision and uncertain accuracy of the literature parallax estimate for GJ 1214 was a limiting factor in our ability to characterize the planet. Significant effort has been made to re-characterize this object, including deriving a different parallax \citep{GJ1214_new} with modern data. For example, if GJ 1214 were located at 15 pc away instead of 13 pc, this would increase the intrinsic luminosity of the star by $0.3$ magnitudes, and increase the inferred radius of the star by 10\%. Clearly, accurately estimating the stellar parameters of M dwarf is vital towards our understanding of any transiting planet they may host. Any subsequent planetary discoveries by the MEarth survey will likely be around a star less well-characterized than GJ 1214, making the challenge of conducting detailed exoplanetary studies even more difficult. 

As MEarth has taken a large number of images of its target stars over the course of entire observing seasons, we investigated whether the MEarth images themselves could be used to measure the trigonometric parallax distances to our targets. We found that the MEarth images are well-suited for astrometric analysis and modified our survey to provide astrometric measurements of all of our stars every $10$ days. This paper presents the results of that effort. In section 2, we present a description of our instruments and our observing strategy. In section 3, we describe our astrometric model, and validate our method with the subset of our sample that have previously measured trigonometric distances. Finally, in section 4 we compare our results to photometric distance estimates, we refine the estimates of the mass and radius of the stars in our survey, and we identify additional stars within 10 parsecs to the Sun.

A few years from now several additional surveys for planets orbiting M-dwarfs will be operational in the northern hemisphere: These include the planned near infrared radial velocity surveys CARMENES  \citep{CARMENES}, the Habitable Planet Finder Spectrograph \citep{HabPlanFin}, and SPIROU \citep{SPIROU}, and the currently operating photometric transit survey APACHE \citep{APACHE}. We hope that the results presented in this paper will be of direct benefit to these projects. At least three additional M-dwarf transit surveys are planned for the southern hemisphere: SPECULOOS \citep{SPECULOOS}, ExTrA (PI: X. Bonfils), and MEarth-South, a copy of the MEarth-North observatory located at the Cerro Tololo Interamerican Observatory in Chile. We may use MEarth-South to undertake a similar effort to provide trigonometric parallaxes for southern targets.

\section{Observations}
The MEarth Observatory is an array consisting of 8 identical f/9 40 cm Ritchey-Chr\'etien telescopes on German equatorial mounts at the Fred Lawrence Whipple Observatory on Mount Hopkins, Arizona. The telescopes are controlled robotically and collect data every clear night from September through July. The facility is closed every August for the summer monsoons. Each telescope contains a 2048 x 2048 pixel CCD with a pixel scale of $\approx0.76"$ / pixel and a Schott RG715 glass filter with anti-reflection coating. The cutoff is defined by the CCD response, and the effective bandpass is similar to the union of the SDSS $i$ and $z$ filters.

Initially we cooled the detectors to $-15^\circ$C and did not use a pre-flash. From 2011 onwards, the MEarth cameras operated at $-30^{\circ}$ C, and before each exposure we adopted a pre-flash of the detector. This increases the dark current (which we subsequently subtract off), but has the benefit of suppressing persistence from the previous exposure. (If the persistent image contained a source that overlaps with the target or reference stars, this could skew our astrometric measurements) There is no discernible effect of any residual persistence on the MEarth photometry, and we do not believe it to be relevant astrometrically. Prior to 2011, we carefully ordered the observing sequence of our target fields to avoid source overlap when switching between them.

We gathered flat-field frames each observing night at dawn and dusk. ÊThe MEarth telescopes use German equatorial mounts, which require the telescope tube to be rotated by 180 degrees relative to the sky upon crossing the meridian. We take advantage of this for flat fielding by obtaining adjacent pairs of flat field images on opposite sides of the meridian to average out large-scale illumination gradients from the Sun and the Moon. ÊOur flat-field correction is further complicated by scattered light concentrating in the center of the field-of-view, where our target is located (the amplitude of this effect is approximately 5\% of the average value across the CCD). ÊConsequently, we filter out all large scale structure from the combined twilight flat field, and use it only to track changes in small scale (high spatial frequency) features such as interpixel sensitivity and dust shadows. ÊThe large scale flat field response was derived from dithered photometry of dense star fields. ÊWe also correct for varying exposure time across the field of view due to shutter travel time, as well as fringing. ÊWe have found these corrections to be stable, and we update them annually.

We measured stellar positions through a method similar to \citet{I85}: Local sky background is estimated by binning each image into 64 pixel $\times$ 64 pixel blocks, and then we estimate the peak of the histogram of the intensity of the pixels within each area. We then interpolate to estimate the background level anywhere in the image from this lower resolution background map \citep{I85}. We measure the stellar location using intensity weighted first moments (also called a centroid), computed over a circular aperture (radius 5 pixels prior to August 2010, and 4 pixels thereafter due to a change in our focus strategy). ÊPixels partially inside the aperture are weighted according to the fraction of the pixel area inside the aperture. ÊThe initial aperture locations are estimated from the expected target location, based on a master MEarth image taken during good weather conditions, and accounting for proper motion. ÊThe solution is then iterated using the measured pixel coordinates to update the location of the aperture. For more details of the MEarth photometric data products and processing pipeline, see \citet{berta2012}.

The MEarth target list consists of $\approx 1800$ nearby M dwarfs selected from \citet{Lepine_33pc_sample}, a subset of the LSPM-North catalog \citep{Lepine_catalog} believed to be within 33 pc of the Sun \citep{Nutzman}. These targets are uniformly distributed across the Northern sky ($\delta > 0^{\circ}$), such that typically only one target exists in the MEarth 26' x 26' field of view, with the exception of multiple systems and occasional unrelated asterisms. Therefore, each field must be observed in a pointed manner, distinguishing MEarth from other photometric transit surveys. Each field of view is, by design, large enough to contain sufficient comparison stars to obtain high precision relative photometry and astrometry. Due to the nature of our targets, the comparison stars are, on average, much bluer (typical $r-J$ = 1.3) than the M-dwarf target (typical $r-J$ = 3.8).

During the summer monsoon, which occurs each year in August, data acquisition was halted and the telescopes were shut down. ÊThis time has been used to perform maintenance activities, and also major upgrades in 2010 and 2011, which may result in significant disturbances to the data, such as changing how the instrument flexes when pointing in different directions. We will describe our procedure for combining pre-2011 and post-2011 data in section 3. ÊAdditional hardware failures have occasionally necessitated removal of the detectors from the telescopes during the observing season, which resulted in rotational offsets (the mechanism for this alignment is repeatable to approximately 0.5 degree). ÊThese changes are likely to result in further disturbances to photometry and astrometry. We describe how we correct for this later. Table \ref{camera_removal} lists the dates where the detectors were removed for each telescope.

From each observing season (September through July), the MEarth project gathered data at a roughly $20-30$ minute cadence for the subset of the targets for which we were actively searching for planets at that time. Beginning in October 2011, we began collecting additional data at a roughly $10$ day cadence for all other targets for the purposes of astrometric measurement. In each pointing, we gathered enough exposures such that we have collected sufficient photons to detect a $2 R_\earth$ planet passing in front of the target star. As a result, different stars will have a different number of exposures per pointing. For the purposes of the astrometry presented here, each individual exposure is treated as a separate data point. The data presented here covers the time period from September 2008 through July 2013.

\section{Analysis}
\subsection{Astrometric Model}
We selected images for inclusion in the astrometry fitting if they met the following criteria: The full-width half-max (FWHM) for the image is less than 5 pixels ($3.5$ arcsec), the average ellipticity of the target stars is less than 0.5, the target star is not more than 15 pixels away from its assigned location on the CCD, and the airmass at which the image was taken is not greater than 2.0. These selection criteria typically eliminate $50\%$ of the MEarth images for each target, but can sometimes eliminate up to $80\%$ of images. Most eliminated frames are eliminated due to a large FWHM, either due to naturally poor seeing or wind-shake of the telescope. For each target star, we selected a master image through an automated routine that selects an image that is of good image quality (low FWHM, ellipticity, sky noise), and good photometric quality (brightness of the stars is not significantly different from a typical exposure). The sky-coordinate system for this image is determined through star matching with the 2MASS catalog. When many images meet the criteria to be a suitable ``master" image, we select one image manually. We investigated the effect of our master frame choice on our final astrometric parameters and find that the choice of master frame does not significantly affect our astrometric measurements. 

Reference stars are selected to be between magnitude 8 and magnitude 13 in the MEarth passband, are not blends or close binaries, and are unambiguously identified in the 2MASS catalog. Additionally, in order to avoid effects due to higher order plate corrections, we attempt to avoid using stars near the edges of the CCD. This is done by initially selecting only stars within 600 pixels of the target star (or in the few cases where there are multiple target stars, the average position of the target stars). If the total number of reference stars within 600 pixels is less than 12, then the radius of the circle is increased by 50 pixels at a time until at least 12 reference stars are selected. We always have at least 12 references stars in each field, and each star is weighted equally in our astrometric analysis.

Our astrometric time series are fit in an iterative manner, first fitting the plate constants for each frame, and then stellar motion parameters for each reference star. This is repeated three times so that the plate constants can converge towards a final solution. These plate constants are then used to fit for the motion parameters of the target star.

Each plate is fit through a least squares method with a six constant linear model based on the positions of the reference stars (the target star is excluded):

\begin{eqnarray}
x'_{i} = A_{i}x_{i} + B_{i}y_{i} + C_{i} \nonumber \\
y'_{i} = D_{i}x_{i} + E_{i}y_{i} + F_{i}
\label{plate_constants_eqn}
\end{eqnarray}

where $x$ and $y$ are the original flux-weighted centroid coordinates of the reference star in pixels, and $A, B, C, D, E,$ and $F$ are the plate constants. $x'$ and $y'$ are the reference star coordinates after the transformation, and are also in pixels. The linear plate constants allow for a different scale in the $x$ and $y$ directions, allow for translation of the frame in both directions, and corrects for any instrument rotation, as well as shearing motion in each frame. When removing the cameras on the MEarth telescopes for repairs or maintenance, disturbances to the data are likely. Our plate constants $[A,B,C,D,E,F]$ remain very close to the identity matrix plus a shift, $[1,0,\Delta x,0,1,\Delta y]$ or a 180 degree rotation plus a shift, $[-1,0,\Delta x,0,-1,\Delta y]$ over all images. The effect of higher order plate constants is mitigated by our selection method for reference stars, and including higher order terms doesn't increase the quality of our fit for stars with previously determined trigonometric parallaxes (see section \ref{bad_model_parameters}). 

Once we have shifted and stretched the frame to align with the master frame, the coordinate system of the master frame (generated through star matching between the master frame and the 2MASS catalog) is applied to convert the $x'$ and $y'$ to $\alpha$ (RA) and $\delta$ (DEC). Then, we fit for both the proper motion for each reference star and the parallax for each reference star through the following method.

First, we remove the proper motion of each reference star since the time of the master frame:

\begin{eqnarray}
\alpha'_{i,s} = \alpha_{i,s} - \mu_{RA,s} \Delta t_i \nonumber \\
\delta'_{i,s} = \delta_{i,s} - \mu_{DEC,s} \Delta t_i
\label{proper_motion_eqn}
\end{eqnarray}

The subscript $i$ denotes each image, $s$ denotes the reference star, and the primed coordinate represents the transformation removing the proper motion since the master image. We then convert the stellar coordinates from RA and DEC to ecliptic longitude ($\lambda$) and ecliptic latitude ($\beta$) through a rotation of the coordinate system. Then, we remove the parallax motion at the image epoch and add the parallax motion from the master image:

\begin{eqnarray}
\lambda_{0,i,s} = \lambda_{i,s} + \pi_{s}(P_{\lambda,0,i,s} - P_{\lambda,i,s}) \nonumber \\
\beta_{0,i,s} = \beta_{i,s} + \pi_{s}(P_{\beta,0,i,s} - P_{\beta,i,s}) 
\label{parallax_eqn}
\end{eqnarray}

where $\pi$ is the parallax amplitude, and $P_{\lambda}$ and $P_{\beta}$ are the parallax factors in each coordinate for each star, $s$ and each image, $i$:

\begin{eqnarray}
P_{\lambda,i,s} = a_i \left( \frac{\sin(\lambda_\sun - \lambda_{0,s})}  {\cos(\beta_0)} \right) \nonumber \\ 
P_{\beta,i,s} = -1.0 \times a_i \left( \frac{\cos(\lambda_\sun - \lambda_{0,s})}  {\sin(\beta_0)} \right)
\end{eqnarray}

where $a_i$ is the Earth-Sun distance in AU at the time of the image, $i$, or the master frame, $0$, and $\lambda_\sun$ is the solar longitude at the time of the image or the master frame. 

Finally, one of eight constants is added to each individual star, for all images, depending on the side of the meridian the image was taken.

\begin{eqnarray}
\lambda_{f,s} = \lambda_{0,s} + G_{s,1,2,3,4} \nonumber \\
\beta_{f,s} = \beta_{0,s} + H_{s,1,2,3,4}
\end{eqnarray}

where the subscripts $1$ and $2$ represent different sides of the meridian for the years 2008-2010 (before we changed the camera housing), while $3$, and $4$ represent the different sides of the meridian from September 2011 onwards. Typical values for G and H are $0.1$ arcsec. 
All data points are weighted equally, and our model is fit using the Levenberg-Marquardt $\chi^2$ method. We note that in order to avoid degeneracies between the parallax and the meridian constants, G and H, it is necessary to obtain data on both sides of the meridian during the same phase of the Earth's orbit, and our data collection strategy was adjusted in the middle of the 2011-2012 observing season in order to resolve this degeneracy. If this degeneracy was present in pre-2011 data for a field of view, then those data are discarded. However, as all observations prior to 2011 were taken at planet-hunting cadence only, this occurrence is very rare, as we observed each field during the entirety of each observing night. Finally, as  the meridian constants are unique to each individual star, we note that the effects of any differential color refraction can be partially corrected for through this method as well.

The plate constants, $A_i$, $B_i$, $C_i$, $D_i$, $E_i$, and $F_i$ (one per image), and the stellar motion parameters, $\mu_{s,RA, DEC}$, $\pi_s$, $G_{s,1,2,3,4}$, and $H_{s,1,2,3,4}$ (one per star) are fit iteratively, while holding the other set fixed, until the solution has converged for the plate constants. If there are more than 12 reference stars available, reference stars whose motions are fit to be $> 1.0$ arcsec yr$^{-1}$ or whose parallax place them closer than $100$ pc are culled from the sample and the plate constants are refit with the remaining stars. We find only 23 reference stars in 17 fields fit these criteria. After these cuts, the median number of reference stars per field is nineteen and the maximum number of reference stars is ninety. 

After the plate constants have converged, we fit the astrometric time series for the target star in the same manner as for the reference stars, fitting for the relative proper motion between the target star and the reference stars, the relative parallax between the target and reference stars, and the meridian constants ($G_{1,2,3,4}$, and $H_{1,2,3,4})$. To avoid the effect of singular outliers, we refit our model after excluding points that lie outside of 3 standard deviations of the residuals for our initial fit. We perform an additional test on our data by refitting the parallax signal while holding the proper motion constant to the value reported in \citet{Lepine_33pc_sample}. If the fit parallax amplitude changes by more than $2\sigma$, then we discard this star from the sample. This test eliminated 41 stars from our sample, and we believe that this is principally due to a degeneracy between the parallax motion and the proper motion, which can be resolved in the future by gathering additional data at the proper phase of the year.

We estimate the internal errors in our trigonometric parallax measurement using a residual permutation algorithm, where we take the residuals from our best fit model, move them over one time stamp, add them back to our best fit model, and then refit the permuted data set for the parallax amplitude, proper motions, as well as our meridian offset constants. We apply this method only to our astrometric time series for the target star and not to the reference stars. This implicitly assumes that the errors in the frame constants (derived from the astrometric precision of the reference stars) are negligible compared to the errors in the astrometric precision of the single target star, which is valid due to the number of reference stars we have contributing to each plate solution. This method, while not allowing the flexibility of generating thousands of fake data sets (except for the minority of our stars that have over a thousand data points), has the benefit of preserving long time-scale correlated noise in our time series, and we find that our derived errors tend to be larger for stars that have fewer total measurements. 

One concern with estimating our errors through this residual permutation algorithm is that for the systems which have less than 50 data points (for which we have 238 out of 1507 objects), there may not exist enough permutations to reliably estimate our error bars. In order to test this, we also estimate our error by refitting 1000 synthetic datasets generated by adding white noise to our observed data with a standard deviation equal to the standard deviation of our initial residuals. We find that the the error bar we find from this method is comparable in magnitude to the residual permutation method, and therefore we elect to quote the error bar derived from the residual permutation method for our results.

\subsection{Correction to Absolute Parallax}
Since our reference stars are all relatively bright stars, the reference frame itself exhibits a small parallax motion, as each reference star is also subject to the observational effects of trigonometric parallax. This effect causes us to systematically measure a smaller parallax angle and therefore a larger distance than if our reference image was static. We used the Besan\c{c}on model of the galaxy \citep{besancon}\footnote{An online tool for generating synthetic star catalogs is graciously provided by the authors at http://model.obs-besancon.fr/} to estimate the parallax of synthetic populations of stars along the same sight-lines as our targets. For each target, we generate a synthetic star catalog oversampled by a factor of $1000$, and select only stars whose apparent magnitudes are between 8.0 and 13.0 in the I-band (an adequate approximation of our non-standard filter for this purpose). We select random subsets of these stars that match our observed reference star magnitude distribution in order to estimate the average distance to a typical reference star and the associated uncertainty. Typical corrections for these stars are between $1$ and $2$ milliarcseconds, but can be as high as $4$ milliarcseconds in certain directions. We note that the uncertainty in the absolute parallax correction is negligible compared to the uncertainty in the relative parallax measurement, and therefore only quote the error in the relative parallax measurement as our total error. 

\subsection{Catalog}
We release here a catalog of our results for each target star. If you want the best estimate for the distance to an individual star, use column 18 and the error in column 16. In order, the columns in this table are:

\begin{enumerate}
\item{\cite{Lepine_catalog} catalog designation number}
\item{Gliese Catalog number}
\item{LHS Catalog number}
\item{NLTT Catalog number}
\item{Right Ascension of object (J2000, hours)}
\item{Right Ascension of object (J2000, minutes)}
\item{Right Ascension of object (J2000, seconds)}
\item{Declination of object (J2000, degrees)}
\item{Declination of object (J2000, minutes)}
\item{Declination of object (J2000, seconds)}
\item{$\mu_{RA} \times$ cos(Dec) from our best fit model}
\item{$\mu_{Dec}$ from our best fit model}
\item{Julian epoch of the first data point in our fit}
\item{Julian epoch of the last data point in our fit}
\item{Relative parallax (milliarcseconds)}
\item{Absolute parallax correction (milliarcseconds)}
\item{Absolute parallax (milliarcseconds)}
\item{Absolute parallax error (milliarcseconds)}
\item{2MASS J magnitude}
\item{2MASS K magnitude}
\item{MEarth estimated mass of the object (using the \citet{Delfosse} relation; see section 4.3)}
\item{MEarth estimated radius of the object (using the Mass-Radius relation presented in equation 10 of \citealt{tabbymr})}
\item{Error in the estimated radius equal to $5\%$ the radius, derived from the scatter in the mass radius relation presented by \citet{tabbymr}}
\item{Number of data points in our fit}
\item{Number of reference stars used in our fit}
\item{MEarth Telescope the data was taken with}
\end{enumerate}

\subsection{Validation using stars with previously determined parallaxes}
\begin{figure}
\centering
\includegraphics[width=0.48\linewidth]{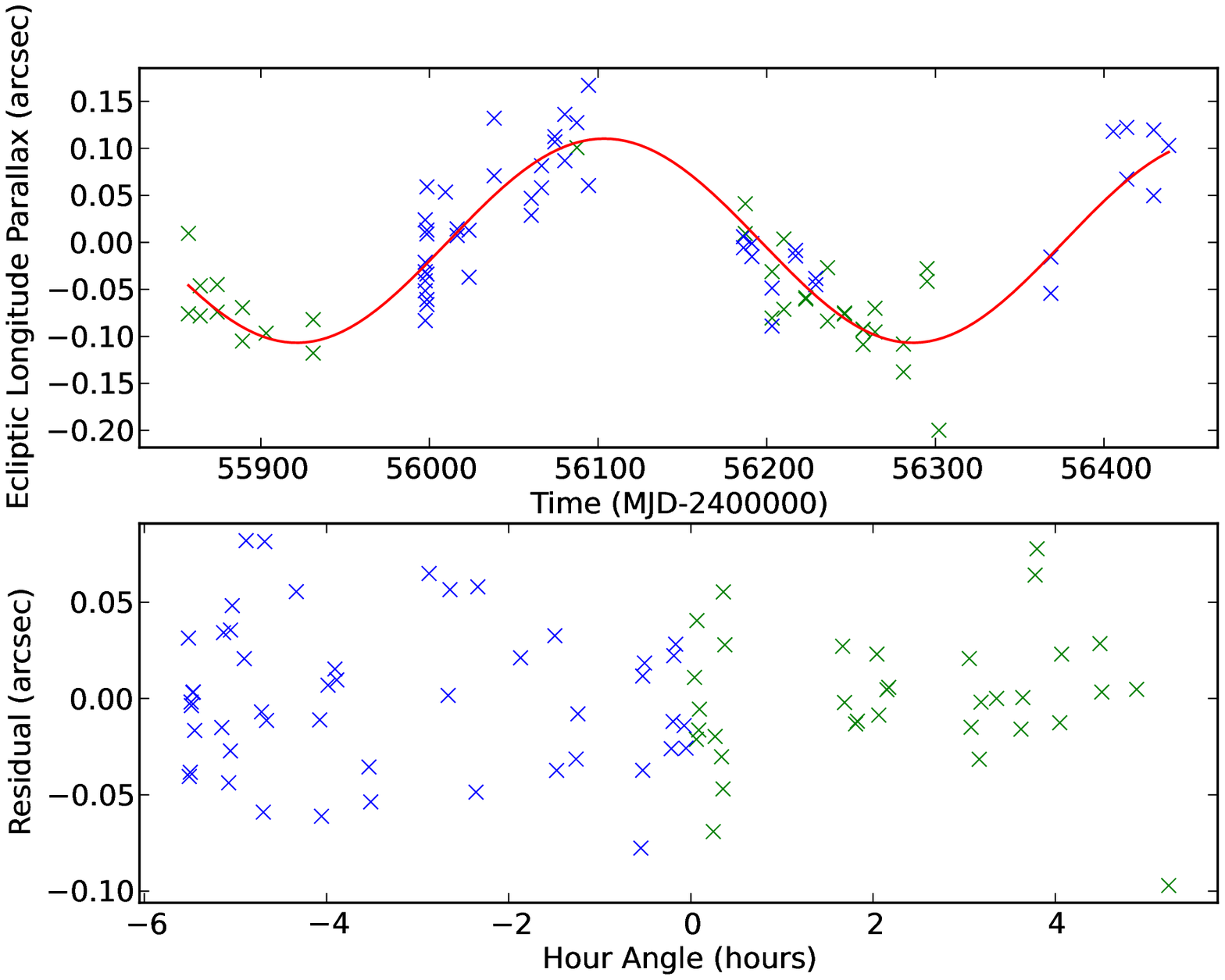} 
\includegraphics[width=0.48\linewidth]{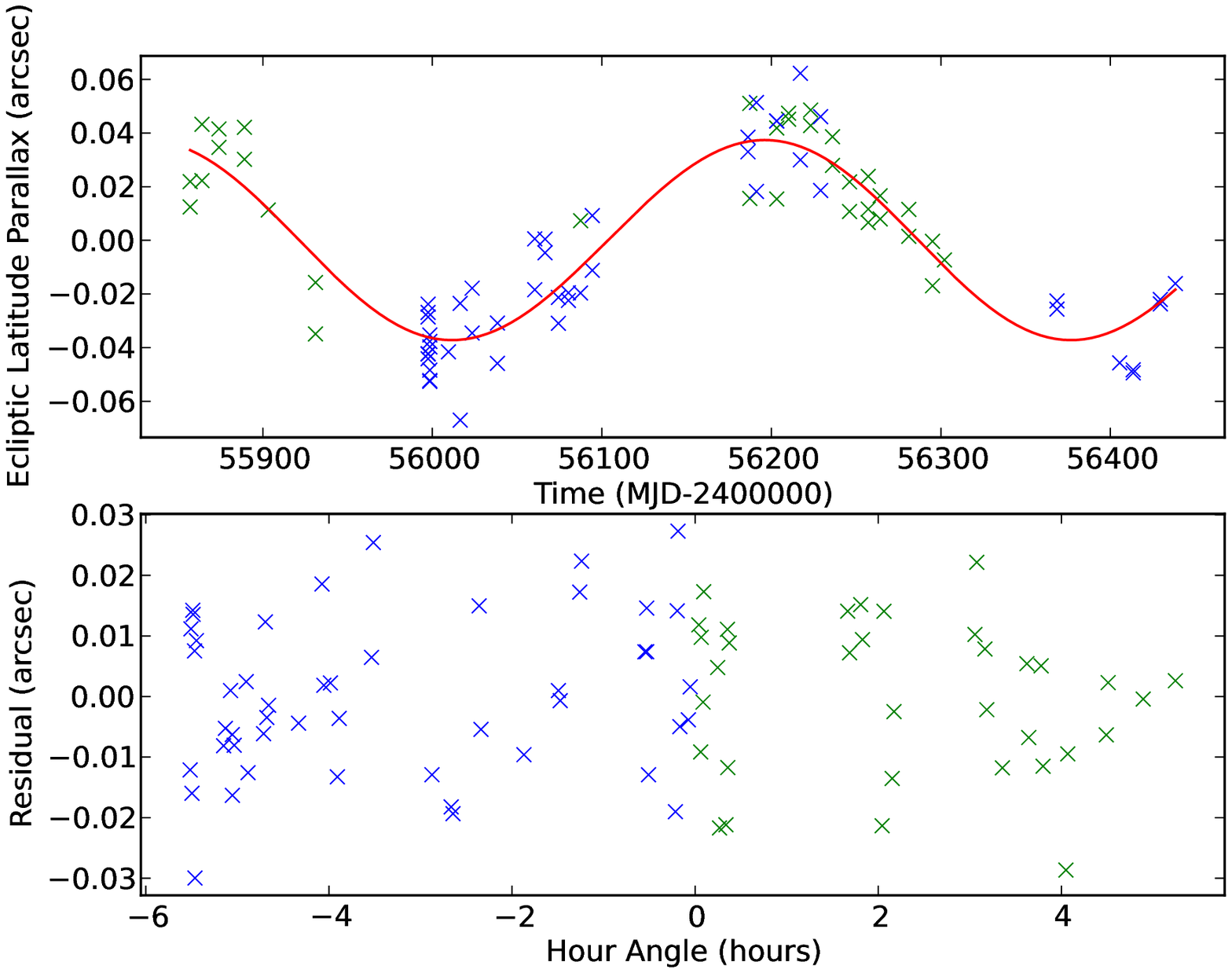}
\caption{Left: Astrometric time series for LSPM J2107+5943 / LHS 64 in ecliptic longitude. Green data points and blue data points represent data taken on opposite sides of the meridian, where MEarth's telescopes have to rotate $180^{\circ}$ due to our German Equatorial mounts. The red line corresponds to the best fit to the MEarth data using our model. Underneath we show the residuals from our best fit model for the ecliptic longitude of LHS 64 versus the hour angle the image was taken.
 Our derived parallax, corrected to absolute parallax of $\pi_{abs} = 41.3 \pm 2.0$ mas is not significantly different from the previous measured $41.8 \pm 2.7$ mas \citep{GJ1222_pi_original}.  %\\
Right: Same as above but in ecliptic latitude instead of ecliptic longitude}
\label{representative}
\end{figure}

\begin{figure}
\centering
\includegraphics[scale=0.6]{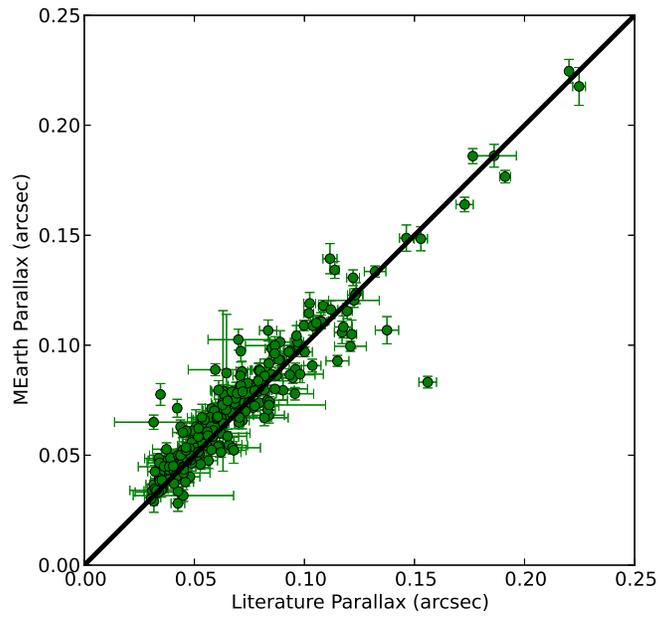} \\
\caption{Comparison of our trigonometric parallaxes derived from the MEarth data with previous studies. We find good agreement between the two measurements, with a typical error of $4$ mas, indicating that MEarth is able to reliably estimate trigonometric parallaxes to our targets. The star in the lower right is LSPM J1631+4051, and we find a trigonometric parallax of $83.2 \pm 2.6$ mas, significantly smaller than the measurement by \citet{Outlier_Reference} of $156 \pm 4$ mas, but much closer to the measurement by \citet{gliese86} of $100 \pm 29$ mas.}
\label{validation_sample}
\end{figure}

\begin{figure}
\centering
\includegraphics[scale=0.6]{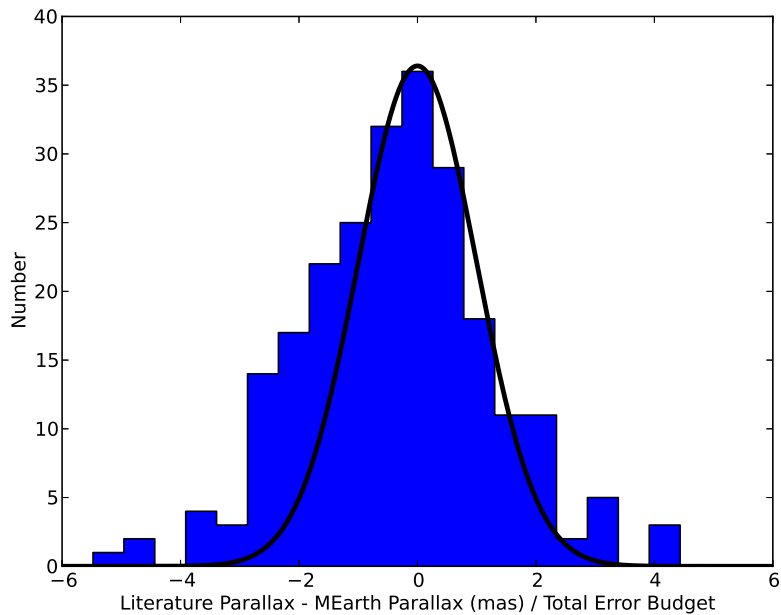} \\
\caption{Ratio of the deviation of our measured parallaxes from previously determined parallaxes compared to the total error budget in the two measurements. The black line is a standard Gaussian with unit variance. We find that our residual permutation algorithm is able to reliably estimate the errors in our measurement, with the distribution being approximately gaussian, with a width only $15\%$ wider than expected. While our error distribution is asymmetric, we note that very few of our measurements are discrepant by more $3\sigma$.}
\label{Error_Budget}
\end{figure}

The subset of the MEarth sample presented here includes 1507 stars for which we can obtain reliable results. This includes 240 stars for which we were able to locate a trigonometric parallax determination in the literature, many of which are from the compilation available in the Yale Parallax Catalog (167 stars, \citealt{YPC}), Hipparcos (41 stars, \citealt{Hipparcos_original}) or \citet{Lepine_33pc_sample}. A representative example of an astrometric time series with MEarth data (for LHS 64) is shown in Figure \ref{representative}. For this star, we find a parallax amplitude of $\pi_{abs} = 0.0412 \pm 0.0017$ arcseconds, not significantly different from the previous determination in \citet{GJ1222_pi_original} of $\pi = 0.0418 \pm 0.0027$ arcseconds, as reported by \citet{YPC}. This data set is taken completely at astrometric cadence, demonstrating that our data collection strategy is sufficient to measure trigonometric parallaxes of our targets. We further note that we measure a parallax for GJ 1214 of $72.8 \pm 2.4$ mas, consistent with the recent determination of $69.1 \pm 0.9$ mas by \citet{GJ1214_new} and the historical determination of $77.2 \pm 5.4$ mas provided by \citet{YPC}. Therefore, we are able to accurately measure apparent stellar motion associated with trigonometric parallax, although in some cases with larger uncertainties than dedicated astrometric programs such as RECONS \citep{RECONS}, the Brown Dwarf Kinematic Project (BDKP, \citealt{BDKP}), and the Solar Neighborhood project (\citealt{solar_neighborhood_24} and references therein). In Figure \ref{validation_sample} we show the MEarth derived parallax motion compared to the values reported in the literature for all stars that had a previous trigonometric parallax measurement available. Literature trigonometric parallax measurements come from \citet{1992AJ....103..638M}, \citet{1993AJ....105.1571H}, \citet{1993PASP..105.1101G},\citet{1995gcts.book.....V}, \citet{YPC}, \citet{1996MNRAS.281..644T}, \citet{Outlier_Reference}, \citet{1999AJ....118.1086B}, \citet{2000AJ....120.1106B}, \citet{2001AJ....121.1607B}, \citet{2002AJ....124.1170D}, \citet{2003AJ....125..354R}, \citet{2004ApJ...617.1323P}, \citet{2006AJ....132.2360H}, \citet{2006ApJ...649..389P}, \citet{2007A&A...464..787S}, \citet{2007A&A...474..653V}, \citet{2007ApJ...668L.155M}, \citet{2008AJ....136..452G}, \citet{2009AJ....137..402G}, \citet{2009AJ....137.4109L}, \citet{2010A&A...514A..84S}, \citet{2010AJ....140..897R}, \citet{2010AstL...36..576K}, \citet{2012ApJ...758...56S}, and \citet{GJ1214_new}. The scatter of the residuals when compared to the reported literature value is approximately Gaussian. Fitting this distribution as a Gaussian, we find a best fit width $\approx15\%$ larger than the sum of our errors and the errors reported in the literature added in quadrature (see Figure \ref{Error_Budget}). 
We report our measured parallaxes and uncertainties for each target in our catalog.

\subsection{Additional Plate Constants, Differential Refraction, and Secular Acceleration}
\label{bad_model_parameters}
To determine whether the plates are modeled adequately by terms that are strictly linear in the coordinates, we refit the subsample of target stars with previously determined trigonometric parallaxes with a 2nd order plate model. We found that the additional parameters ended up improving our derived value of the parallax relative to the previously determined value for only approximately half of the stars, and the remainder of the stars showed a marginal decrease in agreement with previous values. Furthermore, on average, the fit became worse, as the standard deviation of the residuals to the literature values increased by $2$ mas. Therefore, we do not believe the lack of higher order terms in our plate model to be a significant source of error in our results, and that including these higher order terms causes us to fit astrometric noise rather than real trends associated with our detectors.

Another possible source of error is differential color refraction (DCR) in the atmosphere through which our measurements are taken. Since the reference stars (from which the plate constants are derived) are, on average, bluer than the target star, the relative effects of DCR between the reference stars and the target star could become important. \citet{Stone_2002} measured the effect of differential color refraction in different optical bands as a function of color and found that the maximum DCR effect in the I-band for stars with a B-V of 2.0 is 12 mas, suggesting that the effect in our bandpass is probably small, as the B-V color of a typical late M-dwarf is approximately 1.8. Nonetheless, we we proceed to estimate the effect that DCR should have on our data.

The effective wavelength of a typical M-dwarf target, accounting for telluric absorption, the filter bandpass, and the detector quantum efficiency is $850$ nm, and is $840$ nm for a typical reference star. We assume a typical target is a M5V star \citep{pickles}, and that a typical reference star is a G2V star. The index of refraction for air at $15^{\circ}$C, and standard atmospheric pressure at these wavelengths is $n_{840\textrm{nm}} = 1.00027482$ and $n_{850\textrm{nm}} = 1.00027477$ \citep{ciddor_dcr}. However, typical conditions at the MEarth observatory on Mt. Hopkins are significantly different than standard atmospheric conditions. The seasonal average temperature is $10^{\circ}$C and the typical pressure is $575$ mm Hg. Additionally, the amount of water vapor in the air also affects index of refraction of the atmosphere. For a typical relative humidity of $30\%$, at these conditions the partial pressure of water is $2.8$ mm Hg \citep{meteorology_today}. Correcting the index of refraction for these effects using the methods in \citet{filippenko_dcr} and \citet{barrell_dcr}, we find that the expected differential color refraction between our reference stars and the target M-dwarfs is typically $6$ milliarcseconds at airmass = 1.41, and 10 milliarcseconds at airmass = 2. 

Our model is capable of accounting for some of the DCR signal with the meridian constant parameters $G$ and $H$. These constants are fit on a star by star basis, and therefore the ``mean" DCR correction term for each star in each dataset becomes merged into the meridian correction for each star. Any residual differential color effect will only be a second order effect and much smaller than $10$ mas, below our threshold for detection.

To investigate whether any significant effects due to airmass or color remain in our data, we show the residuals in each coordinate direction (ecliptic longitude and ecliptic latitude) as a function of the hour angle at which the image was taken, in Figure \ref{representative}. We find no significant directional offset between the residuals and the airmass the image was taken, and conclude that the effect of differential color refraction in the MEarth bandpass on our astrometry is negligible. 

For nearby stars with large radial velocity, the effect of secular acceleration (a changing of a star's angular proper motion as a result of its changing solar distance) may become important, as our model assumes a constant angular velocity for each star. Barnard's star, one of the fastest moving stars in the sky, has a secular acceleration of approximately 1.2 mas / yr \citep{sec_acc}. Since our astrometric model is fitting for the average proper motion over a maximum of a 4 year time period, this can only result in a maximum systematic of $\approx 2.5$ mas, below our detection threshold. Therefore, we ignore any effects of secular acceleration for all of our targets. 

We further investigated whether the residuals in our derived parallaxes compared to previous results was correlated with other external parameters. We find no correlation with the brightness of our target star, the average brightness of our reference stars, the color of the target star, the average color of the reference stars, or with intrapixel variation (evaluated by looking at our residuals as a function of sub-pixel position).

\section{Results and Discussion}
\subsection{Comparison to Photometric Distance Estimates}
\begin{figure}
\centering
\includegraphics[scale=0.6]{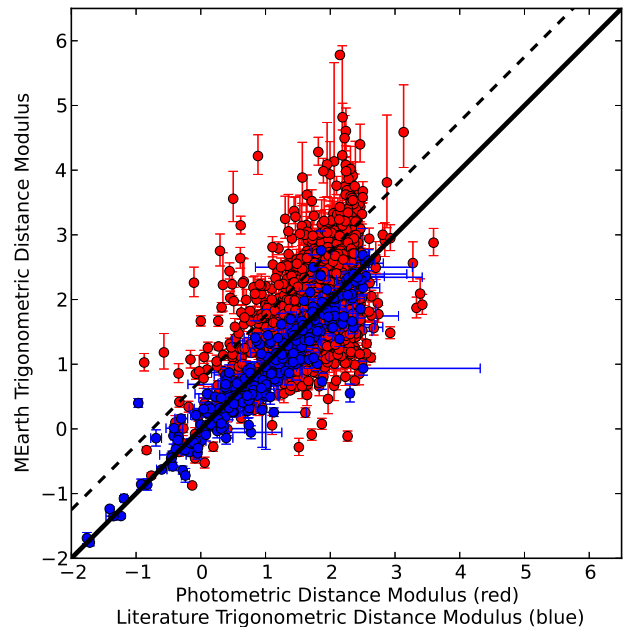} \\
\caption{Plot of the MEarth distance modulus versus the expected distance modulus from either previous literature measurements (blue) or the piece-wise linear V-J fit from \citet{Lepine_33pc_sample} (red). Unresolved equal mass binaries should fall on the dashed line. We note that the MEarth astrometric result is consistent with previous parallax studies and has a much lower dispersion than photometric estimates.}
\label{photometric_distances}
\end{figure}

All of the stars in our sample have estimated distances from the \citet{Lepine_33pc_sample} piece-wise linear relationship in $V_{est}-J$ color. The $V_{est}$-magnitudes for our targets all come from this catalog. This relation was calibrated from the 3104 M dwarfs from the LSPM North catalog \citep{Lepine_catalog} that had trigonometric parallax measurements, and has a mean error of $35\%$ on the distance estimate. We note that most of the estimated $V$ magnitudes in \citet{Lepine_33pc_sample}  come from photographic plate measurements, and some have uncertainties as large as $0.5$ magnitudes. We note that high quality V photometry is available for some of the stars where photographic estimates were used by \citet{Lepine_33pc_sample}, however for the purposes of this work, we use the estimated V magnitudes compiled by \citet{Lepine_33pc_sample}. In Figure \ref{photometric_distances}, we show the distance modulus as expected from the photometric distance, using the calibration from \citet{Lepine_33pc_sample}, compared to the value derived from the MEarth astrometry, as well as stars that have previous trigonometric parallax determinations available. In this plot, an equal mass binary would have an offset of 0.75 in distance modulus from the photometric distance modulus estimate. However, we find that that photometric distance estimates have a significantly higher scatter than trigonometric measurements, and that the typical scatter in the photometric measurement is large enough that identifying equal mass binaries through comparison of photometric estimates with our trigonometric parallaxes is not trivial. 

We note that previous estimates of the binary fraction among M0-M5V stars place the binarity fraction at $42\% \pm 9\%$ \citep{mdwarf_binarity_fischer}, $27\% \pm 16\%$ \citep{mdwarf_binarity_hyades}, or $27\% \pm 3\%$ \citep{astralux_binarity}, with a trend towards smaller binarity fraction for lower mass primaries (which the MEarth survey preferentially targets). \citet{mdwarf_binarity_fischer} find that the binary distribution for M-dwarfs peaks with companion object having an orbital period between 9 and 220 years, and they used a range of detection techniques, including spectroscopy, speckle interferometry, and direct imaging. \citet{mdwarf_binarity_hyades} and \citet{astralux_binarity} relied on imaging, and lucky-imaging techniques respectively, and therefore are not as sensitive to tighter multiple systems. Regardless, there is likely to be a significant amount of contamination with target stars that are actually unresolved binaries. As the MEarth survey target list was designed with a distance cut-off, and it is harder to resolve close binaries at larger distances, it is likely that the contamination is due primarily to unresolved binary systems further than 33 pc masquerading as single stars estimated to be within 33pc. Using only a photometric distance measurement means that the volume in which an equal mass binary can masquerade as a single star is $2^{3/2}$ larger than the volume in which we aim to obtain our sample. Removing these stars from our sample before investing a significant amount of observing time to investigate whether these stars have transiting planets will make the MEarth survey more effective at finding planets. With significantly improved stellar distances, the limiting factor in distinguishing unresolved multiples from single stars is the quality of the available photometry that can be either compiled or gathered for these objects. We are currently working on calibrating the MEarth data to obtain accurate absolute optical magnitudes, and hope in the near future to use this data to determine which of our targets are likely binaries. However, for the rest of this paper we assume that all stars in our sample are single stars.

\subsection{Survey Completeness and Mapping the Solar Neighborhood}
\begin{figure}
\centering
\includegraphics[scale=0.6]{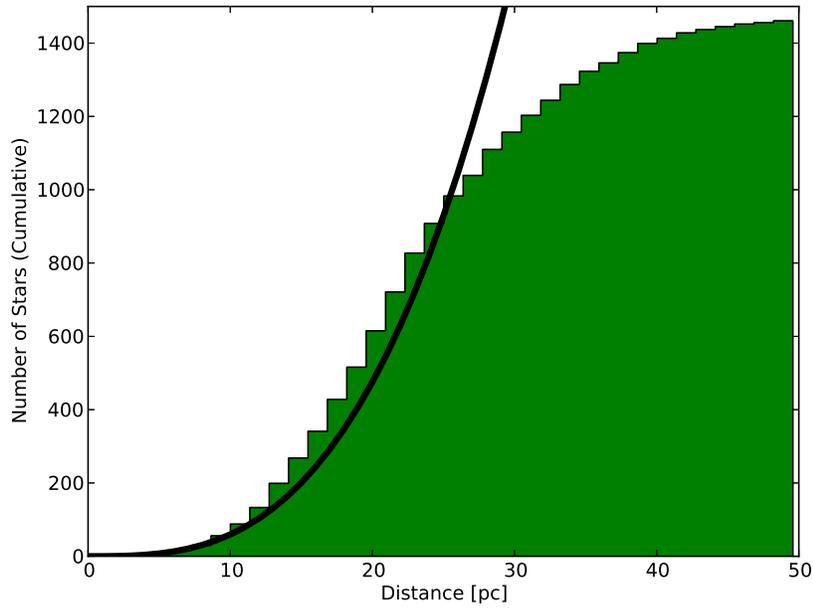} \\
\caption{Top: Cumulative histogram of the number of MEarth targets as a function of distance from the Sun. Overplotted is a black $R^3$ line, normalized to the best fit cubic for the number counts between 5 and 15 parsecs.} %\\
\label{survey_completeness}
\end{figure}

The MEarth survey is designed to look at nearby mid-to-late M dwarfs within 33 pc to the Sun, primarily drawn from the catalog compiled by \citet{Lepine_33pc_sample}. However, we note that in designing this sample, we have introduced spectral type dependent metallicity biases, as the $V_{est}$ magnitude has a dependence on metallicity as well as spectral type. The \citet{Lepine_33pc_sample} catalog was designed to be approximately $50\%$ complete out to 33 pc with the goal of being mostly complete out to a distance of 25 pc. With our estimates of the trigonometric parallaxes for each star, we are now in a position to determine how complete the census of the solar neighborhood is out to these distances for the types of star in the MEarth sample. Assuming that on these small scales the effects of galactic structure are negligible, we can assume that the cumulative number of MEarth targets should increase linearly with the volume, with radius R as $R^3$. In Figure \ref{survey_completeness}, we show the cumulative number of MEarth targets out to 50 pc, with a best fit $R^3$ line, fit to the cumulative number of stars from 5pc to 15 pc. We omit the cumulative number of stars on scales smaller than 5 pc because there are not a sufficient number of stars to make a reliable estimate.

\begin{figure}
\centering
\includegraphics[scale=0.55]{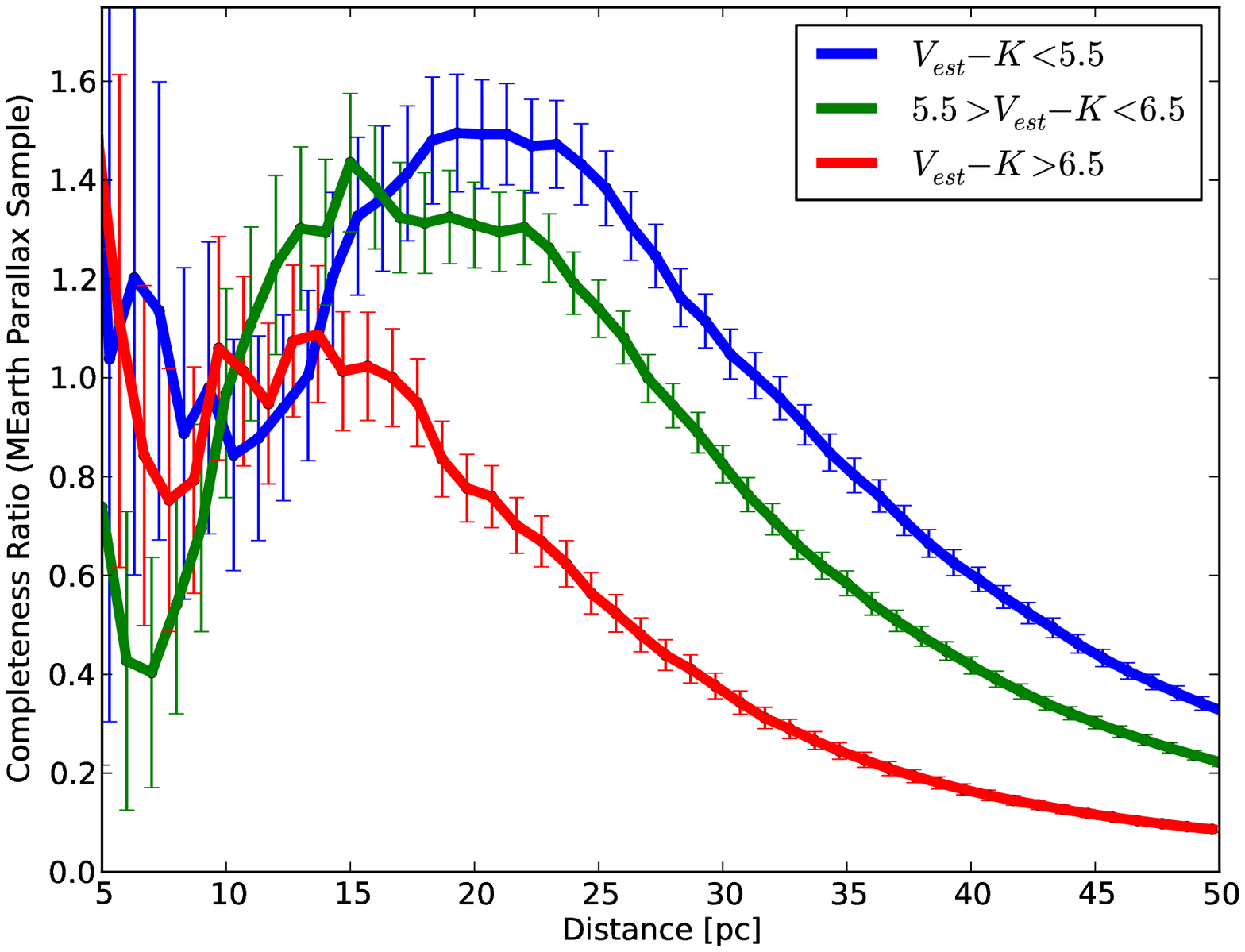} \\
\includegraphics[scale=0.55]{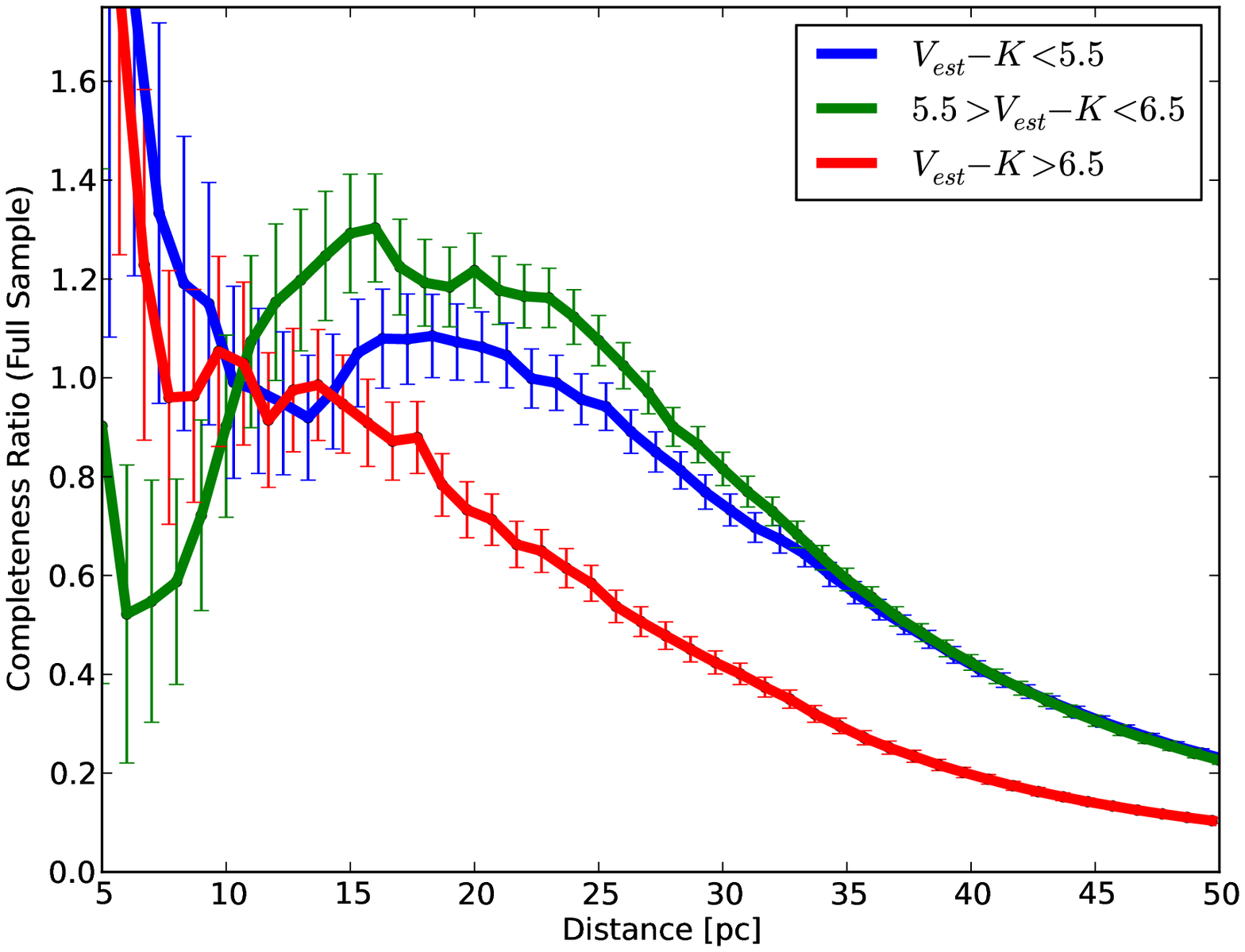}
\caption{Top: Completeness ratio for stars with $V_{est}-K < 5.5$ (approximately M4.5 and earlier, blue), $6.5 > V_{est}-K > 5.5$ (between M4.5 and M5.5V, green), and $V_{est}-K > 6.5$ (later than M5.5V, red), where the completeness ratio is defined as the cumulative number of stars within a distance limit, over what would be expected from a cubic function fit to the cumulative number between 5 and 15 parsecs. 
Bottom: Same as above, but for the full sample of stars from \citet{Nutzman}, where we use the trignometric distances presented here when applicable, and photometric or spectroscopic distances from \citet{Lepine_33pc_sample} for the remainder of the stars.}
\label{completeness_ratios}
\end{figure}

\begin{figure}
\centering
\includegraphics[scale=0.6]{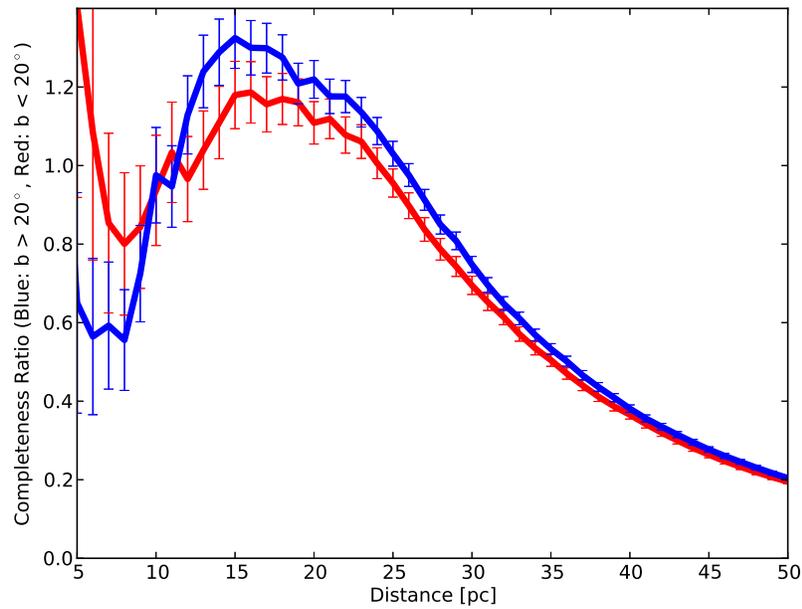} \\
\caption{Completeness ratio for stars with galactic lattitude $b > 20$ degrees (blue) and $b < 20$ degrees (red), where the completeness ratio is defined as the number of stars detected, over what would be expected from a cubic function fit to the cumulative number between 5 and 15 parsecs. We find no correlation with galactic coordinate and completeness, indicating that confusion and source crowding is not the limiting factor in identifying nearby M-dwarfs. Error bars are strictly Poisson.}
\label{completeness_ratio_galactic}
\end{figure}

In Figure \ref{completeness_ratios}, we plot the ratio of the number of stars to the number we expect to find for stars with a $V_{est}-K < 5.5$ (M4.5 and earlier), $5.5 < V_{est}-K < 6.5$ (between M4.5 and M5.5V), and for stars with a $V_{est}-K > 6.5$ (spectral types later than M5.5V). We plot both the sample of stars for which we measure parallaxes (presented here), and all stars that originally made the initial MEarth target selection from \citet{Nutzman}, for which the majority MEarth has taken data. For stars that do not have a measured parallax from the MEarth data, we use an estimated distance from a spectroscopic measurement (if available) or its $V-K$ color. Each color bin is fit to an $R^{3}$ power law using the cumulative number counts between 5 and 15 parsecs, and the ratio of the number of stars we find within that distance to that expected from our fit is what we define as the completeness ratio. 
The sample of new MEarth parallaxes used to construct this plot has one known incompleteness: it is missing some stars earlier than M4.5 in spectral type at distances less than 15 pc, because these stars were too bright for MEarth to observe efficiently. However, because we base our completeness estimates on the number of stars between 5 and 15 pc, we therefore underpredict the local density of M4.5 dwarfs and earlier and appear overcomplete at distances above 15 pc for these stars in the top panel of Figure \ref{completeness_ratios}. As expected, this artifact largely disappears in the bottom panel of Figure \ref{completeness_ratios}, when we include all stars regardless of whether MEarth measured a new parallax for them.
For the full sample, we find that for stars with a $V_{est}-K$ color less than 6.5, we are nearly complete out to a distance of 25 parsecs, but for redder stars, our completeness begins dropping off at the smaller distance of $20$ parsecs. New searches for these smaller, redder objects, utilizing WISE data for example, may identify these missing objects in the near future. We do not find a clear correlation between completeness and Galactic latitude, which indicates crowding and confusion in the Galactic plane may not be the limiting factor in identifying the missing systems (see Figure \ref{completeness_ratio_galactic}).

\subsection{Changes to Stellar Physical Parameters and Application to the MEarth Planet Survey}
We can now estimate the masses and radii of our stars more reliably, aiding our understanding of the physical characteristics of individual M-dwarfs in the solar neighborhood. Specifically, we can more reliably estimate the intrinsic brightness of our targets, and through previously published relationships, estimate the mass and radius of each star. 

\citet{Delfosse} obtained an empirical relation between the masses of low mass stars and their luminosities in the near infrared, with the smallest scatter obtained in the $K$ band. As each of the MEarth targets have precise $K$ band magnitudes as measures by 2MASS \citep{2mass}, we can use these measurements with our trigonometric parallax measurements to obtain more reliable masses for each star. Similarly, by using the mass-radius relation for low mass stars obtained by \citet{Mass_Radius_Relation}, we can obtain more reliable radii for each star. Both of these are important for characterizing MEarth's sensitivity to planetary transits in these systems. 

\begin{figure}
\centering
\includegraphics[scale=0.55]{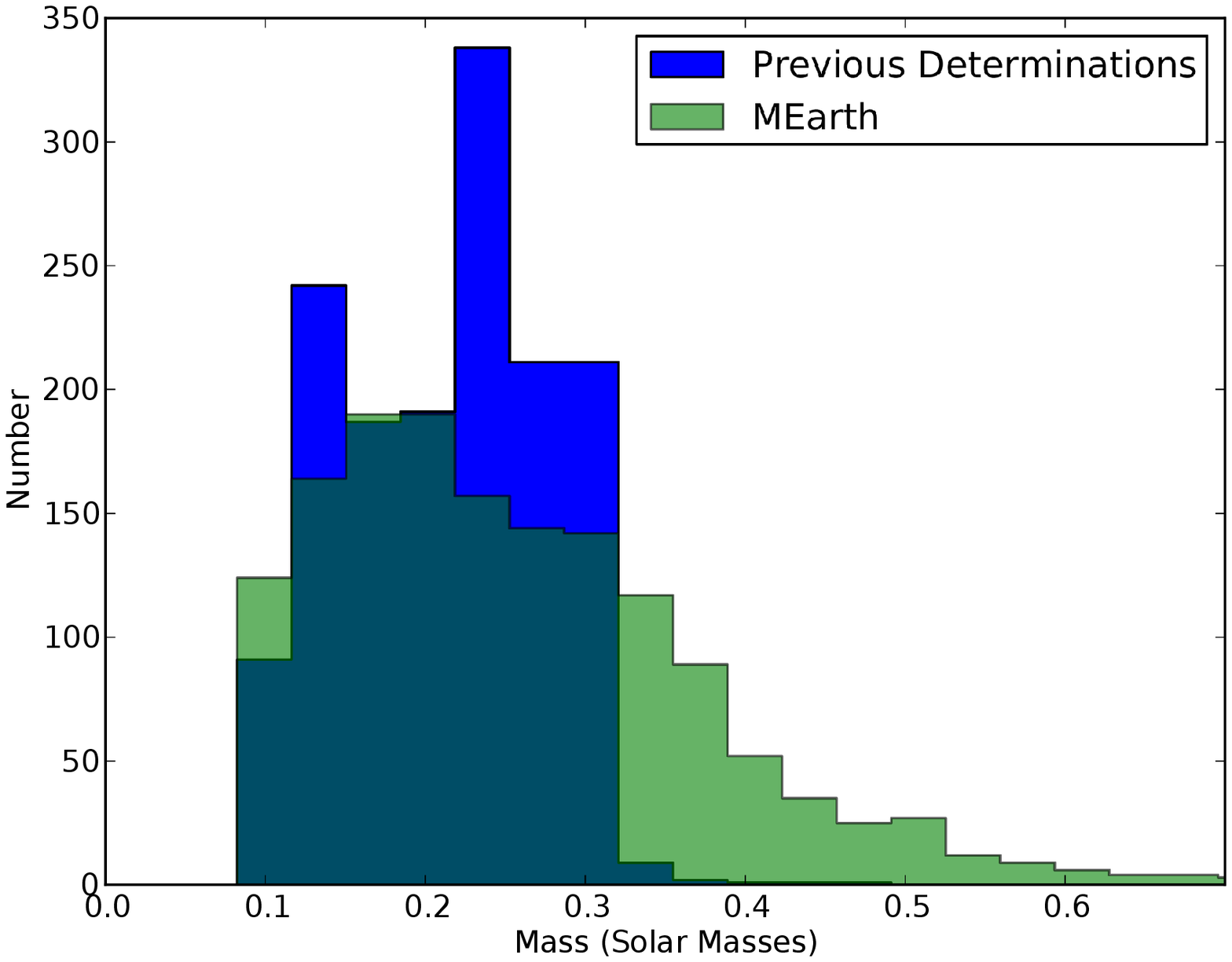} \\
\includegraphics[scale=0.55]{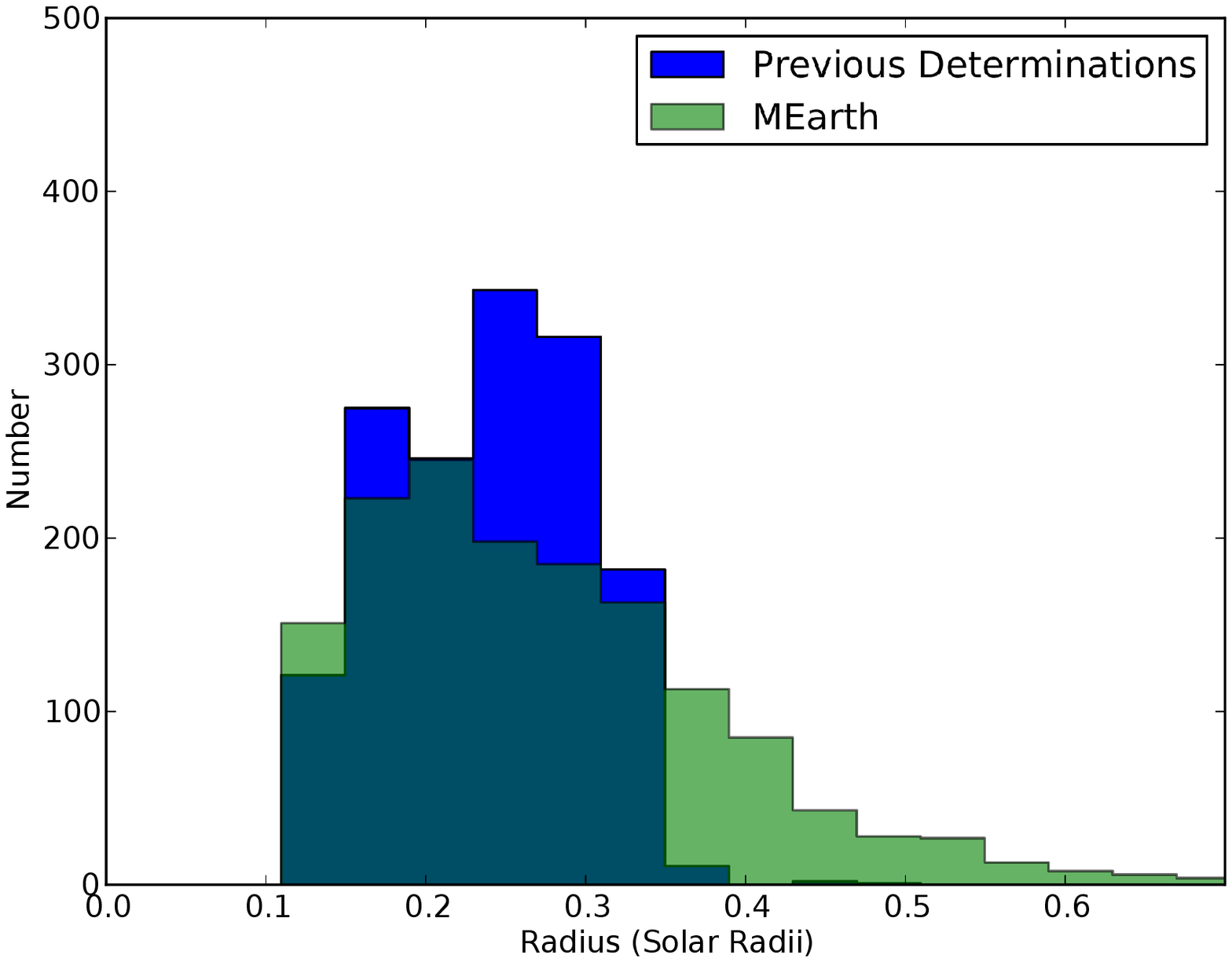}
\caption{Top: Histogram of the distribution of stellar masses of the MEarth target stars based on our previous photometric determinations (blue, \citet{Nutzman}) and based on the absolute K magnitude derived from the MEarth astrometric data and the 2MASS survey (green). We find the peak of our distribution shifts towards smaller mass stars but there is a significant long tail of stars with higher masses (due to being further away) than previously estimated. We note that at the extremely high mass end ($M >\sim 0.75 M_\sun$), the \citet{Delfosse} relation is no longer accurate, and that stars in this region are likely to be unresolved binaries, as color information makes them unlikely to be earlier type stars. Bottom: Identical, but transforming mass to radius with the relation published by \citet{Mass_Radius_Relation}.}
\label{stellar_parameters}
\end{figure}

In Figure \ref{stellar_parameters}, we show histograms of our newly derived stellar masses and radii compared to the values derived by \citet{Nutzman}, which used the photometric distance measurement and the same mass-radius relationship. For stars where the newly estimated mass is $M < 0.5 M_\sun$, we find a median of the absolute value of the offset of $\Delta M \approx 0.08 M_\sun$, whereas if we take the global sample, we find a median of $\Delta M \approx 0.12 M_\sun$, as the initial MEarth sample was constructed to be limited to only the mid-to-late M dwarfs. We note that the stars that we assign higher mass values are located further away than photometrically indicated and may instead be unresolved binary or multiple systems. Additionally, a systematic trend of the $V_{est} - J$ colors being too red would also systematically shift our stars to higher estimated mass once we have obtained trigonometric parallax distances to them. Similarly, the typical change in the stellar radius is also $\Delta R \approx 0.08 R_\sun$ for stars whose mass from MEarth astrometry was $M < 0.5 M_\sun$ and $\Delta R \approx 0.12 R_\sun$ for stars whose final mass was $M < 0.8 M_\sun$

Importantly, the stellar radii of our targets determine our sensitivity towards transiting exoplanets in the system, as the transit depth is approximately proportional to the ratio of the areas of the stellar disk and planetary disk. Since MEarth's observing strategy is to obtain a per-pointing signal to noise ratio sufficient to detect a $2 R_\earth$ planet transiting in front of the star at a significance of $3\sigma$, an accurate estimation of the radius is essential to our ability to detect a transiting planet of a given size. Since large planets tend to be rare around small stars \citep{M_dwarf_planet_distribution_0, M_dwarf_planet_distribution_1, Dressing}, obtaining the appropriate exposure time is essential for MEarth to accomplish its science goals. Additionally, the mass of the host star directly determines the planet's orbital period at a given separation, and directly affects the temperature of the planet as well. Therefore, characterization of a planet directly depends on the host star's physical parameters. As of the 2012-2013 observing season, MEarth is using these new radius determinations to determine which targets are observed in a given season and to set the integration times for these targets.

Aside from the transit depth of a detected companion being dependent on the stellar radius, we also note that obtaining accurate stellar parameters will also affect our understanding of any discovered planet's habitability. The boundary of habitable zone is currently a topic of debate (see, for example, \citealt{hzone1, hzone2, hzone3}), but even in the most simple definition (distance at which the equilibrium temperature of a blackbody supports the existence of liquid water), the period of a planet in the habitable zone of the more massive M-dwarfs in our sample have longer orbital periods than what can be easily detected with the MEarth observatory. Therefore, by shifting our priority to smaller M-dwarfs, we are sensitive to smaller planets at all orbital distances, and increase our rate of recovery for planets in the habitable zone of the target star.

\subsection{New Stars Within 10 pc of the Sun}
While the census of the solar neighborhood out to 10 parsecs is largely complete down to the mid-M dwarfs, there are several stars for which photometric distances places them within or near to the 10 parsec boundary, but for which there is no published trigonometric parallax. In the sample presented here, we find 37 stars whose trigonometric parallaxes place them within the 10 parsec radius boundary. Of those 37 stars, 29 of them have previously measured parallaxes that confirm this distance. The rest the stars have photometric $V-K$ distance estimates that place them within 10 parsecs of the Sun, and we therefore confirm their proximity to the Sun.

\section{Conclusions}
With trigonometric parallaxes for $1507$ stars, we have greatly enhanced our map of the solar neighborhood. When the MEarth survey concludes, we will have also probed this sample with nightly cadence over a multi-month baseline, to measure rotation periods, activity levels, flare occurrences, and the presence of small planetary bodies approximately $2.0 R_\earth$ in orbits extending into the classical habitable zone of these stars (P $\approx 14$ days for an M5V star). The scientific yield from the MEarth survey extends far beyond the goal of finding planets, and in particular, new trigonometric parallaxes can already begin to yield additional insights into stellar astrophysics - particularly the unclear relationship between luminosity, metallicity, age, and activity. 

We have presented here a catalog of trigonometric parallaxes for 1507 mid-to-late M-dwarfs, 1267 of which had no previous trigonometric parallax measurement. The revised
distances indicate 8 stars without previous trigonometric parallax
measurements lie within 10pc of the Sun. We have increased the number of stars in the MEarth sample with trigonometric distance estimates by a factor of 6, and increased the total number of stars within 25 pc of the Sun with trigonometric parallax measurements by 682 stars, $\approx 30\%$ of the currently identified $25$ parsec sample. 

Furthermore, our study has better informed us of the stellar parameters of our targets, most importantly mass and radius. This knowledge has changed the way we have conducted our survey by changing the priority with which we survey our stars. Accurate stellar characterization is an essential step in any transiting planet survey, and these measurements will greatly enhance the MEarth survey, and reduce our uncertainty in the planet occurrence rate around mid to late M-dwarfs. In addition, MEarth-South, a copy of the current MEarth-North array, will be operational in Chile in the near future, surveying an additional $\approx$ 2000 stars in the Southern hemisphere. We will begin operating this array with astrometric and planet-hunting cadence as well, and expect to obtain accurate trigonometric parallaxes for these stars as early as $1$ year after first light, with fits improving in subsequent years.

Our efforts to characterize the nearby stars in the solar neighborhood complement the GAIA mission, which is set to launch at the end of 2013. Over the course of its mission, GAIA will systematically survey the entire sky, obtaining astrometric precision of several microarcseconds for the brightest stars in the sky and collecting data down to a limiting magnitude of $V = 20$. However, it will take several years for GAIA to collect enough data to disentangle the effects of proper motion and parallax motion, while current ongoing photometric and RV surveys for planets around small stars need accurate data to characterize their systems in the short term. Until more precise data is obtained by the GAIA spacecraft, the catalog presented in this paper can characterize these stellar systems.

\newpage
\acknowledgments
We would like to thank the input of our referee, Dr. John Subasavage, whose comments have improved the quality of this manuscript.
The authors would like to acknowledge W. Olson for his extensive assistance in code optimization, without which our results would have taken far longer to compile. We acknowledge the work of M Irwin, which we used in this research.
We gratefully acknowledge funding for the MEarth Project from the David and Lucile Packard Fellowship for Science and Engineering and from the National Science Foundation (NSF) under grant numbers AST-0807690 and AST-1109468. The MEarth team is grateful to the staff at the Fred Lawrence Whipple Observatory for their efforts in construction and maintenance of the facility and would like to thank W. Peters, T. Groner, K. Erdman-Myres, G. Alegria, R. Harris, B. Hutchins, D. Martina, D. Jankovsky, T. Welsh, R. Hyne, M. Calkins, P. Berlind, and G. Esquerdo for their support. This research has made use of NASA's Astrophysics Data System.

{\it Facilities:} \facility{FLWO: MEarth} \\

\bibliography{bibliography}

\begin{thebibliography}{95}
\expandafter\ifx\csname natexlab\endcsname\relax\def\natexlab#1{#1}\fi

\bibitem[{{Ahrens}(1994)}]{meteorology_today}
{Ahrens}, D. 1994, {Meteorology Today - an introduction to weather, climate and
  the environment} (West Publishing Co. 5th ed.)

\bibitem[{{Anglada-Escud{\'e}} {et~al.}(2013){Anglada-Escud{\'e}},
  {Rojas-Ayala}, {Boss}, {Weinberger}, \& {Lloyd}}]{GJ1214_new}
{Anglada-Escud{\'e}}, G., {Rojas-Ayala}, B., {Boss}, A.~P., {Weinberger},
  A.~J., \& {Lloyd}, J.~P. 2013, \aap, 551, A48

\bibitem[{{Barrell}(1951)}]{barrell_dcr}
{Barrell}, H. 1951, J. Opt. Soc. Am., 41, 295

\bibitem[{{Bayless} \& {Orosz}(2006)}]{Mass_Radius_Relation}
{Bayless}, A.~J., \& {Orosz}, J.~A. 2006, \apj, 651, 1155

\bibitem[{{Bean} {et~al.}(2011){Bean}, {D{\'e}sert}, {Kabath}, {Stalder},
  {Seager}, {Miller-Ricci Kempton}, {Berta}, {Homeier}, {Walsh}, \&
  {Seifahrt}}]{gj1214_bean}
{Bean}, J.~L., {D{\'e}sert}, J.-M., {Kabath}, P.,  {et~al.} 2011, \apj, 743, 92

\bibitem[{{Benedict} {et~al.}(1999){Benedict}, {McArthur}, {Chappell}, {Nelan},
  {Jefferys}, {van Altena}, {Lee}, {Cornell}, {Shelus}, {Hemenway}, {Franz},
  {Wasserman}, {Duncombe}, {Story}, {Whipple}, \&
  {Fredrick}}]{1999AJ....118.1086B}
{Benedict}, G.~F., {McArthur}, B., {Chappell}, D.~W.,  {et~al.} 1999, \aj, 118,
  1086

\bibitem[{{Benedict} {et~al.}(2001){Benedict}, {McArthur}, {Franz},
  {Wasserman}, {Henry}, {Takato}, {Strateva}, {Crawford}, {Ianna}, {McCarthy},
  {Nelan}, {Jefferys}, {van Altena}, {Shelus}, {Hemenway}, {Duncombe}, {Story},
  {Whipple}, {Bradley}, \& {Fredrick}}]{2001AJ....121.1607B}
{Benedict}, G.~F., {McArthur}, B.~E., {Franz}, O.~G.,  {et~al.} 2001, \aj, 121,
  1607

\bibitem[{{Benedict} {et~al.}(2000){Benedict}, {McArthur}, {Franz},
  {Wasserman}, \& {Henry}}]{2000AJ....120.1106B}
{Benedict}, G.~F., {McArthur}, B.~E., {Franz}, O.~G., {Wasserman}, L.~H., \&
  {Henry}, T.~J. 2000, \aj, 120, 1106

\bibitem[{{Berta} {et~al.}(2012{\natexlab{a}}){Berta}, {Charbonneau},
  {D{\'e}sert}, {Miller-Ricci Kempton}, {McCullough}, {Burke}, {Fortney},
  {Irwin}, {Nutzman}, \& {Homeier}}]{gj1214_hubble}
{Berta}, Z.~K., {Charbonneau}, D., {D{\'e}sert}, J.-M.,  {et~al.}
  2012{\natexlab{a}}, \apj, 747, 35

\bibitem[{{Berta} {et~al.}(2012{\natexlab{b}}){Berta}, {Irwin}, {Charbonneau},
  {Burke}, \& {Falco}}]{berta2012}
{Berta}, Z.~K., {Irwin}, J., {Charbonneau}, D., {Burke}, C.~J., \& {Falco},
  E.~E. 2012{\natexlab{b}}, \aj, 144, 145

\bibitem[{{Bessel}(1838)}]{bessel}
{Bessel}, F.~W. 1838, Astronomische Nachrichten, 16, 65

\bibitem[{{Bigourdan}(1909)}]{french_catalog}
{Bigourdan}, G. 1909, Bulletin Astronomique, Serie I, 26, 466

\bibitem[{{Boyajian} {et~al.}(2012){Boyajian}, {von Braun}, {van Belle},
  {McAlister}, {ten Brummelaar}, {Kane}, {Muirhead}, {Jones}, {White},
  {Schaefer}, {Ciardi}, {Henry}, {L{\'o}pez-Morales}, {Ridgway}, {Gies}, {Jao},
  {Rojas-Ayala}, {Parks}, {Sturmann}, {Sturmann}, {Turner}, {Farrington},
  {Goldfinger}, \& {Berger}}]{tabbymr}
{Boyajian}, T.~S., {von Braun}, K., {van Belle}, G.,  {et~al.} 2012, \apj, 757,
  112

\bibitem[{{Charbonneau} {et~al.}(2009){Charbonneau}, {Berta}, {Irwin}, {Burke},
  {Nutzman}, {Buchhave}, {Lovis}, {Bonfils}, {Latham}, {Udry}, {Murray-Clay},
  {Holman}, {Falco}, {Winn}, {Queloz}, {Pepe}, {Mayor}, {Delfosse}, \&
  {Forveille}}]{gj1214b}
{Charbonneau}, D., {Berta}, Z.~K., {Irwin}, J.,  {et~al.} 2009, \nat, 462, 891

\bibitem[{{Ciddor}(1996)}]{ciddor_dcr}
{Ciddor}, P.~E. 1996, Applied Optics, 35, 1566

\bibitem[{{Croll} {et~al.}(2011){Croll}, {Albert}, {Jayawardhana},
  {Miller-Ricci Kempton}, {Fortney}, {Murray}, \& {Neilson}}]{gj1214_croll}
{Croll}, B., {Albert}, L., {Jayawardhana}, R.,  {et~al.} 2011, \apj, 736, 78

\bibitem[{{Crossfield} {et~al.}(2011){Crossfield}, {Barman}, \&
  {Hansen}}]{Crossfield_gj1214_spectrum}
{Crossfield}, I.~J.~M., {Barman}, T., \& {Hansen}, B.~M.~S. 2011, \apj, 736,
  132

\bibitem[{{Dahn} {et~al.}(2002){Dahn}, {Harris}, {Vrba}, {Guetter}, {Canzian},
  {Henden}, {Levine}, {Luginbuhl}, {Monet}, {Monet}, {Pier}, {Stone}, {Walker},
  {Burgasser}, {Gizis}, {Kirkpatrick}, {Liebert}, \&
  {Reid}}]{2002AJ....124.1170D}
{Dahn}, C.~C., {Harris}, H.~C., {Vrba}, F.~J.,  {et~al.} 2002, \aj, 124, 1170

\bibitem[{{Danchi} \& {Lopez}(2013)}]{hzone2}
{Danchi}, W.~C., \& {Lopez}, B. 2013, \apj, 769, 27

\bibitem[{{de Mooij} {et~al.}(2013){de Mooij}, {Brogi}, {de Kok}, {Snellen},
  {Croll}, {Jayawardhana}, {Hoekstra}, {Otten}, {Bekkers}, {Haffert}, \& {van
  Houdt}}]{gj1214_rayleigh}
{de Mooij}, E.~J.~W., {Brogi}, M., {de Kok}, R.~J.,  {et~al.} 2013, \apj, 771,
  109

\bibitem[{{Delfosse} {et~al.}(2000){Delfosse}, {Forveille}, {S{\'e}gransan},
  {Beuzit}, {Udry}, {Perrier}, \& {Mayor}}]{Delfosse}
{Delfosse}, X., {Forveille}, T., {S{\'e}gransan}, D.,  {et~al.} 2000, \aap,
  364, 217

\bibitem[{{D{\'e}sert} {et~al.}(2011){D{\'e}sert}, {Bean}, {Miller-Ricci
  Kempton}, {Berta}, {Charbonneau}, {Irwin}, {Fortney}, {Burke}, \&
  {Nutzman}}]{Desert_atmo}
{D{\'e}sert}, J.-M., {Bean}, J., {Miller-Ricci Kempton}, E.,  {et~al.} 2011,
  \apjl, 731, L40

\bibitem[{{Dressing} \& {Charbonneau}(2013)}]{Dressing}
{Dressing}, C.~D., \& {Charbonneau}, D. 2013, \apj, 767, 95

\bibitem[{{Ducourant} {et~al.}(1998){Ducourant}, {Dauphole}, {Rapaport},
  {Colin}, \& {Geffert}}]{Outlier_Reference}
{Ducourant}, C., {Dauphole}, B., {Rapaport}, M., {Colin}, J., \& {Geffert}, M.
  1998, \aap, 333, 882

\bibitem[{{Dupuy} \& {Liu}(2012)}]{trent}
{Dupuy}, T.~J., \& {Liu}, M.~C. 2012, \apjs, 201, 19

\bibitem[{{Faherty} {et~al.}(2012){Faherty}, {Burgasser}, {Walter}, {Van der
  Bliek}, {Shara}, {Cruz}, {West}, {Vrba}, \& {Anglada-Escud{\'e}}}]{BDKP}
{Faherty}, J.~K., {Burgasser}, A.~J., {Walter}, F.~M.,  {et~al.} 2012, \apj,
  752, 56

\bibitem[{{Filippenko}(1982)}]{filippenko_dcr}
{Filippenko}, A.~V. 1982, \pasp, 94, 715

\bibitem[{{Fischer} \& {Marcy}(1992)}]{mdwarf_binarity_fischer}
{Fischer}, D.~A., \& {Marcy}, G.~W. 1992, \apj, 396, 178

\bibitem[{{Fraine} {et~al.}(2013){Fraine}, {Deming}, {Gillon}, {Jehin},
  {Demory}, {Benneke}, {Seager}, {Lewis}, {Knutson}, \& {D{\'e}sert}}]{Fraine}
{Fraine}, J.~D., {Deming}, D., {Gillon}, M.,  {et~al.} 2013, \apj, 765, 127

\bibitem[{{Gaidos}(2013)}]{hzone1}
{Gaidos}, E. 2013, \apj, 770, 90

\bibitem[{{Gaidos} {et~al.}(2012){Gaidos}, {Fischer}, {Mann}, \&
  {L{\'e}pine}}]{M_dwarf_planet_distribution_0}
{Gaidos}, E., {Fischer}, D.~A., {Mann}, A.~W., \& {L{\'e}pine}, S. 2012, \apj,
  746, 36

\bibitem[{{Gatewood}(2008)}]{2008AJ....136..452G}
{Gatewood}, G. 2008, \aj, 136, 452

\bibitem[{{Gatewood} \& {Coban}(2009)}]{2009AJ....137..402G}
{Gatewood}, G., \& {Coban}, L. 2009, \aj, 137, 402

\bibitem[{{Gatewood} {et~al.}(1993){Gatewood}, {de Jonge}, \&
  {Stephenson}}]{1993PASP..105.1101G}
{Gatewood}, G., {de Jonge}, K.~J., \& {Stephenson}, B. 1993, \pasp, 105, 1101

\bibitem[{{Giclas} {et~al.}(1971){Giclas}, {Burnham}, \& {Thomas}}]{giclas_71}
{Giclas}, H.~L., {Burnham}, R., \& {Thomas}, N.~G. 1971, {Lowell proper motion
  survey Northern Hemisphere. The G numbered stars. 8991 stars fainter than
  magnitude 8 with motions $>$ 0''.26/year} (Flagstaff, Arizona: Lowell
  Observatory, 1971)

\bibitem[{{Giclas} {et~al.}(1978){Giclas}, {Burnham}, \& {Thomas}}]{giclas_78}
{Giclas}, H.~L., {Burnham}, Jr., R., \& {Thomas}, N.~G. 1978, Lowell
  Observatory Bulletin, 8, 89

\bibitem[{{Gillon} {et~al.}(2013){Gillon}, {Jehin}, {Delrez}, {Magain},
  {Opitom}, \& {Sohy}}]{SPECULOOS}
{Gillon}, M., {Jehin}, E., {Delrez}, L.,  {et~al.} 2013, in Protostars and
  Planets VI, Heidelberg, July 15-20, 2013. Poster \#2K066, 66

\bibitem[{{Gizis} \& {Reid}(1995)}]{mdwarf_binarity_hyades}
{Gizis}, J., \& {Reid}, I.~N. 1995, \aj, 110, 1248

\bibitem[{{Gliese} {et~al.}(1986){Gliese}, {Jahreiss}, \& {Upgren}}]{gliese86}
{Gliese}, W., {Jahreiss}, H., \& {Upgren}, A.~R. 1986, in The Galaxy and the
  Solar System, ed. R.~{Smoluchowski}, J.~M. {Bahcall}, \& M.~S. {Matthews},
  13--34

\bibitem[{{Harrington} \& {Dahn}(1980)}]{GJ1222_pi_original}
{Harrington}, R.~S., \& {Dahn}, C.~C. 1980, \aj, 85, 454

\bibitem[{{Harrington} {et~al.}(1993){Harrington}, {Dahn}, {Kallarakal},
  {Guetter}, {Riepe}, {Walker}, {Pier}, {Vrba}, {Luginbuhl}, {Harris}, \&
  {Ables}}]{1993AJ....105.1571H}
{Harrington}, R.~S., {Dahn}, C.~C., {Kallarakal}, V.~V.,  {et~al.} 1993, \aj,
  105, 1571

\bibitem[{{Henry} {et~al.}(2006{\natexlab{a}}){Henry}, {Jao}, {Subasavage},
  {Beaulieu}, {Ianna}, {Costa}, \& {M{\'e}ndez}}]{RECONS}
{Henry}, T.~J., {Jao}, W.-C., {Subasavage}, J.~P.,  {et~al.}
  2006{\natexlab{a}}, \aj, 132, 2360

\bibitem[{{Henry} {et~al.}(2006{\natexlab{b}}){Henry}, {Jao}, {Subasavage},
  {Beaulieu}, {Ianna}, {Costa}, \& {M{\'e}ndez}}]{2006AJ....132.2360H}
---. 2006{\natexlab{b}}, \aj, 132, 2360

\bibitem[{{Irwin}(1985)}]{I85}
{Irwin}, M.~J. 1985, \mnras, 214, 575

\bibitem[{{Janson} {et~al.}(2012){Janson}, {Hormuth}, {Bergfors}, {Brandner},
  {Hippler}, {Daemgen}, {Kudryavtseva}, {Schmalzl}, {Schnupp}, \&
  {Henning}}]{astralux_binarity}
{Janson}, M., {Hormuth}, F., {Bergfors}, C.,  {et~al.} 2012, \apj, 754, 44

\bibitem[{{Jao} {et~al.}(2011){Jao}, {Henry}, {Subasavage}, {Winters},
  {Riedel}, \& {Ianna}}]{solar_neighborhood_24}
{Jao}, W.-C., {Henry}, T.~J., {Subasavage}, J.~P.,  {et~al.} 2011, \aj, 141,
  117

\bibitem[{{Jenkins}(1952)}]{1952_cat}
{Jenkins}, L.~F. 1952, {General catalogue of trigonometric stellar parallaxes.}
  ([New Haven, Yale University Observatory] 1952.)

\bibitem[{{Kapteyn}(1900)}]{Kapteyn1900}
{Kapteyn}, J.~C. 1900, Publications of the Kapteyn Astronomical Laboratory
  Groningen, 1, 3

\bibitem[{{Khrutskaya} {et~al.}(2010){Khrutskaya}, {Izmailov}, \&
  {Khovrichev}}]{2010AstL...36..576K}
{Khrutskaya}, E.~V., {Izmailov}, I.~S., \& {Khovrichev}, M.~Y. 2010, Astronomy
  Letters, 36, 576

\bibitem[{{Kopparapu} {et~al.}(2013){Kopparapu}, {Ramirez}, {Kasting}, {Eymet},
  {Robinson}, {Mahadevan}, {Terrien}, {Domagal-Goldman}, {Meadows}, \&
  {Deshpande}}]{hzone3}
{Kopparapu}, R.~K., {Ramirez}, R., {Kasting}, J.~F.,  {et~al.} 2013, \apj, 765,
  131

\bibitem[{{L{\'e}pine}(2005)}]{Lepine_33pc_sample}
{L{\'e}pine}, S. 2005, \aj, 130, 1680

\bibitem[{{L{\'e}pine} \& {Gaidos}(2011)}]{LandG11}
{L{\'e}pine}, S., \& {Gaidos}, E. 2011, \aj, 142, 138

\bibitem[{{L{\'e}pine} \& {Shara}(2005)}]{Lepine_catalog}
{L{\'e}pine}, S., \& {Shara}, M.~M. 2005, \aj, 129, 1483

\bibitem[{{L{\'e}pine} {et~al.}(2009){L{\'e}pine}, {Thorstensen}, {Shara}, \&
  {Rich}}]{2009AJ....137.4109L}
{L{\'e}pine}, S., {Thorstensen}, J.~R., {Shara}, M.~M., \& {Rich}, R.~M. 2009,
  \aj, 137, 4109

\bibitem[{{Luyten}(1979)}]{nltt_catalog}
{Luyten}, W.~J. 1979, {New Luyten catalogue of stars with proper motions larger
  than two tenths of an arcsecond; and first supplement; NLTT. (Minneapolis
  (1979))}

\bibitem[{{Mahadevan} {et~al.}(2012){Mahadevan}, {Ramsey}, {Bender}, {Terrien},
  {Wright}, {Halverson}, {Hearty}, {Nelson}, {Burton}, {Redman}, {Osterman},
  {Diddams}, {Kasting}, {Endl}, \& {Deshpande}}]{HabPlanFin}
{Mahadevan}, S., {Ramsey}, L., {Bender}, C.,  {et~al.} 2012, in Society of
  Photo-Optical Instrumentation Engineers (SPIE) Conference Series, Vol. 8446,
  Society of Photo-Optical Instrumentation Engineers (SPIE) Conference Series

\bibitem[{{Makarov} {et~al.}(2007){Makarov}, {Zacharias}, {Hennessy}, {Harris},
  \& {Monet}}]{2007ApJ...668L.155M}
{Makarov}, V.~V., {Zacharias}, N., {Hennessy}, G.~S., {Harris}, H.~C., \&
  {Monet}, A.~K.~B. 2007, \apjl, 668, L155

\bibitem[{{Monet} {et~al.}(1992){Monet}, {Dahn}, {Vrba}, {Harris}, {Pier},
  {Luginbuhl}, \& {Ables}}]{1992AJ....103..638M}
{Monet}, D.~G., {Dahn}, C.~C., {Vrba}, F.~J.,  {et~al.} 1992, \aj, 103, 638

\bibitem[{{Morton} \& {Swift}(2013)}]{M_dwarf_planet_distribution_1}
{Morton}, T.~D., \& {Swift}, J.~J. 2013, submitted, ArXiv:1303.3013

\bibitem[{{Murgas} {et~al.}(2012){Murgas}, {Pall{\'e}}, {Cabrera-Lavers},
  {Col{\'o}n}, {Mart{\'{\i}}n}, \& {Parviainen}}]{Murgas}
{Murgas}, F., {Pall{\'e}}, E., {Cabrera-Lavers}, A.,  {et~al.} 2012, \aap, 544,
  A41

\bibitem[{{Nutzman} \& {Charbonneau}(2008)}]{Nutzman}
{Nutzman}, P., \& {Charbonneau}, D. 2008, \pasp, 120, 317

\bibitem[{{Perryman} {et~al.}(1997){Perryman}, {Lindegren}, {Kovalevsky},
  {Hoeg}, {Bastian}, {Bernacca}, {Cr{\'e}z{\'e}}, {Donati}, {Grenon},
  {Grewing}, {van Leeuwen}, {van der Marel}, {Mignard}, {Murray}, {Le Poole},
  {Schrijver}, {Turon}, {Arenou}, {Froeschl{\'e}}, \&
  {Petersen}}]{Hipparcos_original}
{Perryman}, M.~A.~C., {Lindegren}, L., {Kovalevsky}, J.,  {et~al.} 1997, \aap,
  323, L49

\bibitem[{{Pickles}(1998)}]{pickles}
{Pickles}, A.~J. 1998, \pasp, 110, 863

\bibitem[{{Pravdo} {et~al.}(2004){Pravdo}, {Shaklan}, {Henry}, \&
  {Benedict}}]{2004ApJ...617.1323P}
{Pravdo}, S.~H., {Shaklan}, S.~B., {Henry}, T., \& {Benedict}, G.~F. 2004,
  \apj, 617, 1323

\bibitem[{{Pravdo} {et~al.}(2006){Pravdo}, {Shaklan}, {Wiktorowicz},
  {Kulkarni}, {Lloyd}, {Martinache}, {Tuthill}, \&
  {Ireland}}]{2006ApJ...649..389P}
{Pravdo}, S.~H., {Shaklan}, S.~B., {Wiktorowicz}, S.~J.,  {et~al.} 2006, \apj,
  649, 389

\bibitem[{{Quirrenbach} {et~al.}(2012){Quirrenbach}, {Amado}, {Seifert},
  {S{\'a}nchez Carrasco}, {Mandel}, {Caballero}, {Mundt}, {Ribas}, {Reiners},
  {Abril}, {Aceituno}, {Alonso-Floriano}, {Ammler-von Eiff}, {Anglada-Escude},
  {Antona Jim{\'e}nez}, {Anwand-Heerwart}, {Barrado y Navascu{\'e}s},
  {Becerril}, {Bejar}, {Benitez}, {Cardenas}, {Claret}, {Colome},
  {Cort{\'e}s-Contreras}, {Czesla}, {del Burgo}, {Doellinger}, {Dorda},
  {Dreizler}, {Feiz}, {Fernandez}, {Galadi}, {Garrido}, {Gonz{\'a}lez
  Hern{\'a}ndez}, {Guardia}, {Guenther}, {de Guindos}, {Guti{\'e}rrez-Soto},
  {Hagen}, {Hatzes}, {Hauschildt}, {Helmling}, {Henning}, {Herrero}, {Huber},
  {Huber}, {Jeffers}, {Joergens}, {de Juan}, {Kehr}, {Klutsch}, {K{\"u}rster},
  {Lalitha}, {Laun}, {Lemke}, {Lenzen}, {Lizon}, {L{\'o}pez del Fresno},
  {L{\'o}pez-Morales}, {L{\'o}pez-Santiago}, {Mall}, {Martin},
  {Mart{\'{\i}}n-Ruiz}, {Mirabet}, {Montes}, {Morales}, {Morales Mu{\~n}oz},
  {Moya}, {Naranjo}, {Oreiro}, {P{\'e}rez Medialdea}, {Pluto}, {Rabaza},
  {Ramon}, {Rebolo}, {Reffert}, {Rhode}, {Rix}, {Rodler}, {Rodr{\'{\i}}guez},
  {Rodr{\'{\i}}guez L{\'o}pez}, {Rodr{\'{\i}}guez P{\'e}rez}, {Rodriguez
  Trinidad}, {Rohloff}, {S{\'a}nchez-Blanco}, {Sanz-Forcada}, {Sch{\"a}fer},
  {Schiller}, {Schmidt}, {Schmitt}, {Solano}, {Stahl}, {Storz}, {St{\"u}rmer},
  {Suarez}, {Thiele}, {Ulbrich}, {Vidal-Dasilva}, {Wagner}, {Winkler}, {Xu},
  {Zapatero Osorio}, \& {Zechmeister}}]{CARMENES}
{Quirrenbach}, A., {Amado}, P.~J., {Seifert}, W.,  {et~al.} 2012, in Society of
  Photo-Optical Instrumentation Engineers (SPIE) Conference Series, Vol. 8446,
  Society of Photo-Optical Instrumentation Engineers (SPIE) Conference Series

\bibitem[{{Reid} {et~al.}(2003){Reid}, {Cruz}, {Laurie}, {Liebert}, {Dahn},
  {Harris}, {Guetter}, {Stone}, {Canzian}, {Luginbuhl}, {Levine}, {Monet}, \&
  {Monet}}]{2003AJ....125..354R}
{Reid}, I.~N., {Cruz}, K.~L., {Laurie}, S.~P.,  {et~al.} 2003, \aj, 125, 354

\bibitem[{{Reshetov} {et~al.}(2012){Reshetov}, {Herriot}, {Thibault},
  {D{\'e}saulniers}, {Saddlemyer}, \& {Loop}}]{SPIROU}
{Reshetov}, V., {Herriot}, G., {Thibault}, S.,  {et~al.} 2012, in Society of
  Photo-Optical Instrumentation Engineers (SPIE) Conference Series, Vol. 8446,
  Society of Photo-Optical Instrumentation Engineers (SPIE) Conference Series

\bibitem[{{Riedel} {et~al.}(2011){Riedel}, {Murphy}, {Henry}, {Melis}, {Jao},
  \& {Subasavage}}]{ctio}
{Riedel}, A.~R., {Murphy}, S.~J., {Henry}, T.~J.,  {et~al.} 2011, \aj, 142, 104

\bibitem[{{Riedel} {et~al.}(2010){Riedel}, {Subasavage}, {Finch}, {Jao},
  {Henry}, {Winters}, {Brown}, {Ianna}, {Costa}, \&
  {Mendez}}]{2010AJ....140..897R}
{Riedel}, A.~R., {Subasavage}, J.~P., {Finch}, C.~T.,  {et~al.} 2010, \aj, 140,
  897

\bibitem[{{Robin} {et~al.}(2003){Robin}, {Reyl{\'e}}, {Derri{\`e}re}, \&
  {Picaud}}]{besancon}
{Robin}, A.~C., {Reyl{\'e}}, C., {Derri{\`e}re}, S., \& {Picaud}, S. 2003,
  \aap, 409, 523

\bibitem[{{Ross}(1939)}]{Ross39}
{Ross}, F.~E. 1939, \aj, 48, 163

\bibitem[{{Schlesinger}(1904)}]{Schlesinger04}
{Schlesinger}, F. 1904, \apj, 20, 123

\bibitem[{{Schlesinger}(1910{\natexlab{a}})}]{Schlesinger_films}
---. 1910{\natexlab{a}}, Publications of the Allegheny Observatory of the
  University of Pittsburgh, 1, 1

\bibitem[{{Schlesinger}(1910{\natexlab{b}})}]{Schlesinger_glass}
---. 1910{\natexlab{b}}, Publications of the Allegheny Observatory of the
  University of Pittsburgh, 1, 101

\bibitem[{{Schlesinger}(1911)}]{Yerkes2}
---. 1911, \apj, 33, 8

\bibitem[{{Schlesinger}(1924)}]{1924_cat}
---. 1924, {Probleme der Astronomie, Seeliger Festschrift} (Springer, Berlin)

\bibitem[{{Schlesinger}(1910{\natexlab{c}})}]{Yerkes1}
{Schlesinger}, I.~F. 1910{\natexlab{c}}, \apj, 32, 372

\bibitem[{{Shkolnik} {et~al.}(2012){Shkolnik}, {Anglada-Escud{\'e}}, {Liu},
  {Bowler}, {Weinberger}, {Boss}, {Reid}, \& {Tamura}}]{2012ApJ...758...56S}
{Shkolnik}, E.~L., {Anglada-Escud{\'e}}, G., {Liu}, M.~C.,  {et~al.} 2012,
  \apj, 758, 56

\bibitem[{{Skrutskie} {et~al.}(2000){Skrutskie}, {Schneider}, {Stiening},
  {Strom}, {Weinberg}, {Beichman}, {Chester}, {Cutri}, {Lonsdale}, {Elias},
  {Elston}, {Capps}, {Carpenter}, {Huchra}, {Liebert}, {Monet}, {Price}, \&
  {Seitzer}}]{2mass}
{Skrutskie}, M.~F., {Schneider}, S.~E., {Stiening}, R.,  {et~al.} 2000, VizieR
  Online Data Catalog, 2241, 0

\bibitem[{{Smart} {et~al.}(2010){Smart}, {Ioannidis}, {Jones}, {Bucciarelli},
  \& {Lattanzi}}]{2010A&A...514A..84S}
{Smart}, R.~L., {Ioannidis}, G., {Jones}, H.~R.~A., {Bucciarelli}, B., \&
  {Lattanzi}, M.~G. 2010, \aap, 514, A84

\bibitem[{{Smart} {et~al.}(2007){Smart}, {Lattanzi}, {Jahrei{\ss}},
  {Bucciarelli}, \& {Massone}}]{2007A&A...464..787S}
{Smart}, R.~L., {Lattanzi}, M.~G., {Jahrei{\ss}}, H., {Bucciarelli}, B., \&
  {Massone}, G. 2007, \aap, 464, 787

\bibitem[{{Sozzetti} {et~al.}(2013){Sozzetti}, {Bernagozzi}, {Bertolini},
  {Calcidese}, {Carbognani}, {Cenadelli}, {Christille}, {Damasso}, {Giacobbe},
  {Lanteri}, {Lattanzi}, \& {Smart}}]{APACHE}
{Sozzetti}, A., {Bernagozzi}, A., {Bertolini}, E.,  {et~al.} 2013, in European
  Physical Journal Web of Conferences, Vol.~47, European Physical Journal Web
  of Conferences, 3006

\bibitem[{{Stone}(2002)}]{Stone_2002}
{Stone}, R.~C. 2002, \pasp, 114, 1070

\bibitem[{{Tinney}(1996)}]{1996MNRAS.281..644T}
{Tinney}, C.~G. 1996, \mnras, 281, 644

\bibitem[{{Upgren}(1996)}]{luyten_obituary}
{Upgren}, A.~R. 1996, \qjras, 37, 453

\bibitem[{{van Altena} \& {Hoffleit}(1996)}]{YPC}
{van Altena}, W., \& {Hoffleit}, D., eds. 1996, {Yale Parallax Catalogue}

\bibitem[{{van Altena} {et~al.}(1995){van Altena}, {Lee}, \&
  {Hoffleit}}]{1995gcts.book.....V}
{van Altena}, W.~F., {Lee}, J.~T., \& {Hoffleit}, E.~D. 1995, {The general
  catalogue of trigonometric [stellar] parallaxes}

\bibitem[{{van de Kamp}(1935)}]{sec_acc}
{van de Kamp}, P. 1935, \aj, 44, 74

\bibitem[{{van Leeuwen}(2007{\natexlab{a}})}]{Hipparcos_reduction_1}
{van Leeuwen}, F., ed. 2007{\natexlab{a}}, Astrophysics and Space Science
  Library, Vol. 350, {Hipparcos, the New Reduction of the Raw Data}

\bibitem[{{van Leeuwen}(2007{\natexlab{b}})}]{Hipparcos_reduction_2}
{van Leeuwen}, F. 2007{\natexlab{b}}, \aap, 474, 653

\bibitem[{{van Leeuwen}(2007{\natexlab{c}})}]{2007A&A...474..653V}
---. 2007{\natexlab{c}}, \aap, 474, 653

\bibitem[{{van Maanen}(1915)}]{van_maanen_1915}
{van Maanen}, A. 1915, \apj, 41, 187

\bibitem[{{Vrba} {et~al.}(2004){Vrba}, {Henden}, {Luginbuhl}, {Guetter},
  {Munn}, {Canzian}, {Burgasser}, {Kirkpatrick}, {Fan}, {Geballe},
  {Golimowski}, {Knapp}, {Leggett}, {Schneider}, \&
  {Brinkmann}}]{USNO_IR_parallax}
{Vrba}, F.~J., {Henden}, A.~A., {Luginbuhl}, C.~B.,  {et~al.} 2004, \aj, 127,
  2948

\bibitem[{{Wolf}(1919)}]{wolf1919}
{Wolf}, M. 1919, Veroeffentlichungen der Badischen Sternwarte zu Heidelberg, 7,
  195

\end{thebibliography}

\clearpage

\begin{table}
\begin{center}
\label{camera_removal}
\caption{Dates cameras were removed from the MEarth telescopes}
\begin{tabular}{rl}
\textit{Telescope} & \textit{UT Date} \\
\hline
1 & 2009 Nov 16 \\
1 & 2010 Oct 29 \\
1 & 2011 Oct 12 \\
1 & 2012 Jan 02 \\
1 & 2012 Feb 01 \\
\hline
2 & 2009 Nov 20 \\
2 & 2010 Oct 29 \\
2 & 2011 Oct 12 \\
2 & 2011 Dec 24 \\
2 & 2012 Sep 04 \\
\hline
3 & 2009 Nov 16 \\
3 & 2010 Oct 29 \\
3 & 2011 Oct 12 \\
3 & 2012 Feb 07 \\
3 & 2012 Jun 22 \\
3 & 2012 Sep 09 \\
3 & 2012 Dec 28 \\
\hline
4 & 2010 Oct 29 \\
4 & 2011 Oct 12 \\
4 & 2012 Apr 26 \\
4 & 2012 Aug 02 \\
\hline
5 & 2009 Nov 16 \\
5 & 2010 Feb 09 \\
5 & 2010 May 26 \\
5 & 2010 Oct 29 \\
5 & 2011 Oct 12 \\
5 & 2012 Jan 08 \\
\hline
6 & 2010 Oct 29 \\
6 & 2011 Oct 12 \\
6 & 2012 Jan 09 \\
\hline
7 & 2009 Nov 16 \\
7 & 2010 Oct 29 \\
7 & 2011 Oct 12 \\
7 & 2012 Apr 18 \\
7 & 2012 Oct 25 \\
7 & 2013 Mar 09 \\
\hline
8 & 2010 Oct 29 \\
8 & 2011 Oct 12 \\
8 & 2012 Jan 02 \\
\end{tabular}
\end{center}
\end{table}

\newpage

\end{document}